\documentclass[twocolumn,epsf,showpacs,prd]{revtex4}
\usepackage{graphicx}
\usepackage{dcolumn}
\usepackage{bm}

\begin{document}

\title{Ultrahigh Energy Nuclei Propagation in a
Structured, Magnetized Universe}

\author{Eric Armengaud$^{a,b}$, G\"unter Sigl$^{a,b}$,
Francesco Miniati$^{c,d}$}

\affiliation{$^a$ F\'ed\'eration de Recherche Astroparticule et
Cosmologie, Coll\`ege de France, 11 place Marcelin Berthelot, 75231 Paris}

\affiliation{$^b$ GReCO, Institut d'Astrophysique de Paris, C.N.R.S.,
98 bis boulevard Arago, F-75014 Paris, France}

\affiliation{$^c$ Physics Department, ETH Z\"urich, 8093 Z\"urich, Switzerland}

\affiliation{$^d$ Max-Planck Institut f\"ur Astrophysik,
Karl-Schwarzschild-Str. 1, 85741 Garching, Germany}

\begin{abstract}
We compare the propagation of iron and proton nuclei above
$10^{19}\,$eV in a structured Universe with source and
magnetic field distributions obtained from a large scale
structure simulation and source densities $\sim 10^{-5}
\mbox{Mpc}^{-3}$. All relevant cosmic ray interactions are taken into account,
including photo-disintegration and propagation of secondary
products. Iron injection predicts
spectral shapes different from proton injection which disagree
with existing data below $\simeq30\,$EeV. Injection of light
nuclei or protons must therefore contribute at these energies.
However, at higher energies, existing data are consistent
with injection
of pure iron with spectral indices between $\sim 2$ and $\sim 2.4$.
This allows a significant recovery of the spectrum above
$\simeq100\,$EeV, especially in the case of large
deflections. Significant auto-correlation and anisotropy,
and considerable cosmic variance are also predicted
in this energy range. The mean atomic mass $A$ fluctuates considerably between
different scenarios. At energies below 60 EeV, if the observed $A\gtrsim35$,
magnetic fields must have a negligible effect on propagation.
At the highest energies the observed flux will be dominated by
only a few sources whose location may be determined by next
generation experiments to within $10-20^\circ$ even if
extra-galactic magnetic fields are important.
\end{abstract}


\maketitle

\section{Introduction}

Ultra-high energy cosmic rays (UHECRs) are particles of energy
$\geq~10^{19}$ eV, which give rise to spectacular air showers
spreading over square miles when reaching the Earth's lower atmosphere,
but whose properties are at the moment poorly known because of low
fluxes scaling roughly as $E^{-3}$. A recent review on the
experimental and theoretical aspects of this topic can
be found, for example, in Ref.~\cite{review}.

As the Auger Observatory~\cite{auger-design-report} is still in
a phase of deployment and early analysis, most of the current
data come from two experiments relying on different techniques, the
High Resolution (HiRes) ``Fly's Eye'' fluorescence telescopes
and the AGASA ground array. The first technique consists in observing
the development of the longitudinal extent of the air shower through
the atmospheric fluorescence yield, whereas arrays reconstruct the
lateral development of the shower, by detecting secondary products
reaching the ground level.

Because of large experimental systematics in energy
determination for both methods, and because of low statistics at
the highest energies,
the flux of UHECRs is still poorly known. Whereas AGASA does not see
any cut-off in the energy spectrum even at $10^{20}$ eV
\cite{agasa-spectrum}, recent results of
HiRes indicate a cut-off above around $10^{19.8}\,$eV~\cite{hires-spectrum}.

The nature of the primary particles is even less clear. AGASA sets an upper
limit on the photon fraction of $\leq 28\%$ at $10^{19}$
eV~\cite{agasa-photons},
but it is experimentally difficult to discriminate between light and
heavy nuclei: various methods give results which are largely dependent
on hadronic models and on experimental uncertainties, leading to
discrepancies in the results~\cite{watson-compo}.

Concerning the arrival directions, all the data accumulated until now
are roughly consistent with isotropy, at least on
large angular scales. However, particularly at the highest energies, the lack
of statistics still allows a substantial large-scale anisotropy of the
UHECR sky. On small scales (a few degrees), AGASA has detected a
clustering of events \cite{agasa-clusters} but its statistical
significance remains questionable \cite{finley}. Furthermore,
the HiRes experiment didn't confirm this feature \cite{hires-clusters},
but the integrated aperture of HiRes stereo data is still lower
than that of AGASA.

The Pierre Auger Observatory which is currently under construction
will within a few years accumulate a statistics on UHECRs at energies
above $10^{19}$ eV which will be orders of magnitude above previous
experiments. The resulting spectrum, arrival direction map and perhaps
composition should therefore put serious constraints on UHECR origin
and propagation models. This provides a motivation for investigating
a wide variety of UHECR scenarios consistent with current data.

Various exotic models for the UHECR origin have been proposed~\cite{bs},
particularly in light of the possible absence of the
Greisen-Zatsepin-Kuzmin (GZK) ``cut-off''~\cite{gzk} as
claimed by AGASA. However, the experimental situation about the
spectrum remains unsettled and classical ``astrophysical'' scenarios remain
completely plausible. These models are based on particle acceleration at
shocks in powerful extragalactic astrophysical objects, ranging from
compact objects
such as $\gamma-$ray bursts (GRBs) to large-scale radio lobes of
active galactic nuclei (AGNs), see for example Ref.~\cite{torres-review}.
They have the obvious advantage to be exclusively based on known physics
and astrophysical objects. They should therefore be tested extensively
before a possible rejection in light of more speculative scenarios.

Much work has already been carried out in the framework of astrophysical
scenarios. Various models for sources and particle propagation predict
different observables such as the spectrum or anisotropies
of UHECRs observed on Earth. A major feature of UHECR propagation
models is the strength and extension of extragalactic magnetic
fields (EGMF) which can give rise to important deflections depending on
particle charge and energy. The EGMF are mostly unknown
at present~\cite{vallee} and different models for these fields can lead to
different predictions, as the comparison between
Refs.~\cite{lss-protons,sme-proc}
and Ref.~\cite{dolag} shows. In Ref.~\cite{lss-protons}, the authors
use magnetic fields derived from a cosmological large scale structure
(LSS) simulation with magnetic fields generated at the shocks that
form during LSS formation, whereas in Ref.~\cite{sme-proc} and
Ref.~\cite{dolag} fields of ``primordial'' origin have been
considered. While the different models for initial magnetic seed fields
produce different large scale magnetic field distributions and,
therefore, lead to different predictions for UHECR deflection,
there is still a significant discrepancy between
Ref.~\cite{lss-protons,sme-proc} and  Ref.~\cite{dolag}, hinting
that other technical reasons may play a role here.

The goal of this article is to extend previous work on proton
propagation in Ref.~\cite{lss-protons} to the case of iron sources. This
allows to fill a gap
in the range of plausible models before Auger data will discriminate between
various possibilities. The Hillas criterion~\cite{hillas} shows
qualitatively that heavy nuclei can be accelerated to higher energies
than protons because their gyroradii in the magnetized accelerator
are smaller. Injection of heavy nuclei can therefore plausibly
contribute to the observed UHECR flux at the highest energies.

Heavy nuclei propagation has already been considered previously~\cite{heavy}:
Discrete sources at given distances were studied with and without
unstructured magnetic fields~\cite{bertone}. A model with discrete
and continuous source distributions and magnetic fields with a
uniform Kolmogorov distribution was developed in Ref.~\cite{yamamoto}.
In Ref.~\cite{sigl-iron}, heavy nuclei propagation was studied
for magnetized individual sources.

In the present paper, we consider the following scenario, largely
inspired by the ``best-fit model'' which was presented in
Ref.~\cite{lss-protons}: The sources are distributed according to
the baryon density obtained from a LSS simulation of a typical
local universe~\cite{miniati}. Their
density, $n_{s} \simeq 2.4\times10^{-5} \mbox{Mpc}^{-3}$, corresponds
to average distances between sources of $\simeq40\,$Mpc, comparable
to typical UHECR interaction lengths at GZK energies. Therefore,
neither the continuous source approximation nor the ``universal
spectrum'' discussed in Ref.~\cite{aloisio} is applicable to this case.
Source densities of this order of magnitude are motivated by
$i$) comparable densities of candidates for powerful accelerators such
as AGNs and $ii$) the fact that they appear necessary to explain the
small-scale clustering observed by AGASA~\cite{kachelriess,lss-protons,blasi}.
The observer is located in a void next to a supercluster of matter,
to mimic our local extragalactic environment. The magnetic fields
there are weak, although an experimental estimate of such a quantity
in our immediate surroundings (i.e. within a few Mpc) is lacking.
However, magnetic fields of the order of a $\mu$G in galaxy
clusters are present and sufficient to significantly deflect high
energy charged particles.

We do not take into account galactic magnetic fields in the work
presented here, although they might play a significant role in the
degradation of anisotropy signals for iron even at super-GZK
energies. Previous studies, e.g. Ref.~\cite{yoshi-galdeflec},
show that galactic and halo fields spread arrival directions
by an angle depending on energy, composition and arrival direction
relative to the galactic center.

The configuration studied, namely rare sources, structured magnetic
fields, and the presence of nuclei at injection, complicates the
simulations. The numerical techniques and difficulties are presented
in Sect.~II. In Sect.~III we turn to the results obtained
with proton injection, and in Sect.~IV we study iron injection in the
case of negligible deflection. Both Sects.~III and IV describe useful
reference cases. The scenario with iron injection including the EGMF
obtained from the LSS simulation of Ref.~\cite{miniati} is developed
and interpreted in Sect.~V. Finally, we conclude in Sect.~VI.

%%%%%%%%%%%%%%%%%%%%%%%%%%%%%%%%%%%%%%%%%%%%%%%%%%%%%%%%%%%%%%%%%%%%%%%%%%

\section{Simulations and Methods}

The numerical framework for our simulations is similar to the one
described in some details in Ref.~\cite{lss-protons}. Therefore we will
here only remind the reader of the major features of the technique, and
present some specific remarks concerning problems raised by the study
of heavy nuclei propagation.

\subsection{Magnetic fields and propagation}

The cosmic environment, i.e. the magnetic field as well as
the baryonic density which defines the source distribution are
the same as described in Ref.~\cite{lss-protons}. They have been
computed according to a simulation of large scale structure formation.
An extensive discussion of the numerical modeling of the magnetic
field was presented already in Ref.~\cite{lss-protons,sme-proc}.
Here it suffices to say that the magnetic fields are generated
at cosmic shocks according to the Biermann battery mechanism
and are then renormalized at the end of the simulation
so that for a Coma-like cluster the average magnetic field in the core
would be of order of a $\mu$G. An alternative mechanism
to generate magnetic fields at shocks is provided by the Weibel
instability~\cite{weibel59}. According to recent investigations of this
process~\cite{silva03,schlick03}, magnetic fields
with strength amounting to
a sizable fraction of the thermal energy can be produced at cosmic shocks
on very short time scales (of order of the inverse of the electron plasma
frequency). In any case, an important feature of our LSS
simulation is the existence of $\sim 0.01-1\mu$G scale EGMF extending over
scales of several Mpc in and around galaxy clusters.
In contrast, EGMF in the voids are $\lesssim10^{-5}\,\mu$G, negligible
for UHECR deflection.
Importantly, the resulting EGMF is consistent with statistics of existing
Faraday Rotation Measures with lines of sight through filaments, despite
the fact that the magnetic field strength can be close to equipartition
value with the total energy of the gas~\cite{ryu}.
We therefore have a strongly structured EGMF,
different from the idealized case of a field with a Kolmogorov
spectrum with spatially constant parameters used in Ref.~\cite{bertone}.

Our EGMF also differ from those resulting
from uniform initial seeds, in that the latter appear to be more
concentrated in the core of collapsed structures~\cite{sme-proc,dolag}.
However, while a more concentrated field produces less deflection
of UHECRs, at least the results from the model in Ref.~\cite{sme-proc}
indicate that such as difference is not dramatic. Different numerical
models predict different results though.
Unfortunately, the behavior of the magnetic field with distance from
the cluster center is not known. The current data indicate that
$\mu$G strong magnetic fields extend out to at least $\sim 1$
Mpc~\cite{clarke}
and possibly to larger distances~\cite{johnston}.
At distances above $1\,$Mpc from a cluster core, however, probing the
magnetic fields becomes extremely difficult because the Faraday
Rotation Measure loses sensitivity in low density
regions. Furthermore, the intracluster magnetic field topology is
also poorly known, although the situation will likely improve
in the future.

The size of the LSS box used in our simulations is $\sim 75\,$Mpc, with
a grid of $5123$ comoving cells. The observer, modeled as a sphere of
1.5 Mpc radius, is placed at the border of a void, not far from
a massive and magnetized structure which mimics
the Virgo cluster. To enhance CPU efficiency and allow at
the same time to have particles reaching the observer from regions
further than 75 Mpc, we use periodic boundary conditions in the
simulation to duplicate the allowed propagation region of UHECRs.

To take into account the ``cosmic variance'' which arises because of
various possible source locations relative to the observer, we simulate
different realizations, for which the position of 10 sources are
chosen at random within the box, with probability proportional
to the baryon density. Iron nuclei or protons are injected isotropically
from each of these random sources in the simulation.

Nuclei interactions are treated as described
in Ref.~\cite{bertone}: We take into account photo-disintegration, pion
production and pair production on the low energy photon backgrounds.
Deflection is computed by solving the
Lorentz equation of motion: For this purpose we use
a Bulirsch-Stoer integrator with adaptive stepsize.

In order to measure properly the all-particle spectrum at the observer
position, we need to keep track of every secondary particle during
the propagation. One nucleus emitted by a source can therefore
generate up to 56 nucleons after a propagation time depending mostly
on primary energy. When recording an ``event'', no distinction is
made between primary and secondary particles, but the properties of the
particle, namely its charge and mass, are recorded.

After propagation, the analysis of simulated events is performed
for each simulated data set, roughly in the same way as in
Ref.~\cite{lss-protons}: Taking into account fluctuations in the
spectra and intensities of the sources, we build observed energy
spectra, average composition plots, deflection
histograms, typical sky maps, auto-correlation functions and
full-sky angular power spectra (partial sky coverage being in theory
invertible as shown in Ref.~\cite{deligny}). 
In the present study we neglect any finite experimental resolution in
energy and arrival directions. The ``cosmic variance''
associated to fluctuations in source positions and properties
within the scenario is also computed for these observables. It is
in general defined as the one-sided median deviation from
the average of the considered quantity. The
fluctuations of the source properties are chosen as in
Ref.~\cite{lss-protons}: Each source is characterized by a spectral
index $\alpha_i$ and a luminosity $Q_i$ with

\begin{eqnarray}
\frac{dn_s}{dQ_i}&\propto& Q_i^{-2.2} \;\;\mbox{for}\;\; 1 \leq Q_i
\leq 100\,,\label{source_fluct}\\
\frac{dn_s}{d\alpha_i}&=&\mbox{const.} \;\;\mbox{for}\;\;
\langle \alpha \rangle-0.1 \leq \alpha_i \leq
\langle \alpha \rangle+0.1\,.\nonumber
\end{eqnarray}

Here, $\langle \alpha \rangle$ will be chosen to best fit the observed
spectrum. Finally, we assume that all sources accelerate to a common
maximal energy $E_{\rm max}$ for which we will choose different
values.

%%%%%%%%%%%%%%%%%%%%%%%%%%%%%%%%%%%%%%%%%%%%%%%%%%%%%%%%%%%%%%%%%%%%%%%%%%

\subsection{Numerical Difficulties}

Our simulations allow detailed simultaneous predictions of various
observables in
specific scenarios. The drawback is of course a large CPU time
consumption: The ratio of simulated trajectories over
recorded events is large, of order 1000 or more depending on the
considered scenario. In the case of nuclei
propagation, and for the EGMF we use, the recorded event
yield is particularly low because of large deflections and the
necessity to follow secondary nuclei.

Unfortunately, backtracing the particles from the
observer does not allow to predict observables for the model, since we do
not know in advance the spectrum, composition, and effective source
distribution of \emph{observed} events, not to mention the stochastic
nature of cosmic ray interactions.

The consequence of CPU limitations is the limited statistics of
simulated events. As we want to simulate many source realizations for
different scenarios (typically around 100 source location
realizations for each scenario), we restricted ourselves to $10^4$ events per
realization, above either 10 EeV or 40 EeV. This is sufficient as a first
step, since our goal is to explore the widest range of possible
scenarios rather than to focus on one specific model, which we might
choose later in the light of Auger results.
In the future, we plan to carry out more detailed simulations using a
parallel version of the cosmic ray propagation code currently under
development.

\subsubsection{Event Reweighting}

For CPU efficiency, we inject particles at the sources with a uniform
distribution in the logarithm of energy. To predict observables with
fluctuating source luminosities $Q_i$ and spectral indexes $\alpha_{i}$,
we reweight each simulated trajectory with a factor that depends on the
source power and the injection spectrum. This reduces the effective number
of events for statistical quantities such as histograms and the
anisotropy observables. Reweighting can also bias the event maps by
causing spots to appear on the maps.

We show here as an example how applying
weights to simulated trajectories reduces the sensitivity of the
angular power spectrum to large-scale anisotropies.

\begin{itemize}
\item{\emph{Absence of weights.}
Given $N$ arrival directions $\vec{n}_i$ distributed on the sphere,
we define our estimator of the $C_{\ell}$ by
$ a_{\ell m}~=~N^{-1} \sum_{i=1}^{N} Y_{\ell m}(\vec{n}_i)$ and
$C_{\ell}~=~(2 \ell +1)^{-1} \sum_{m= - \ell}^{\ell} | a_{\ell m} |^2$,
where $Y_{\ell m}(\vec{n}_i)$ are the usual spherical harmonics.
For the null hypothesis of full isotropy, the mean value for this
estimator becomes

$$ \left< C_{\ell} \right> = \frac{1}{(2\ell +1)\,N^2} \sum_{m}
\sum_{i,j} \left< Y_{\ell m}(\vec{n}_i) Y_{\ell m}^{*}(\vec{n}_j)
\right>\,.$$

As the arrival directions are independent, only the
terms $i = j$ contribute, for which
$ \left< | Y_{\ell m}(\vec{n}_i) |^2 \right>
= (4\pi)^{-1}\int d\Omega  | Y_{\ell m}(\vec{n}) |^2 = (4\pi)^{-1}$,
and therefore, one finds the well-known result

\begin{equation}
\left< C_{\ell}\right> = \frac{1}{4\pi N}\,.\label{c1}
\end{equation}

\item{\emph{Effect of weights:} Let us now assign weights $\omega_i$
to the events, with a distribution $p(\omega)$ with mean $\mu$ and
standard deviation $\sigma$. The first average to consider is over
arrival directions. The computation is the same as before, but since
the definition of the estimator is now $a_{\ell m}~=~\left(\sum_j
\omega_j\right)^{-1}\sum_i \omega_i Y_{\ell m}(\vec{n}_i)$,
it follows that $\left< C_{\ell} \right>~=~(4\pi)^{-1}\left(\sum
\omega_i^2\right)\left(\sum \omega_i\right)^{-2}$.
The next step is to average over the weight distribution, which is
independent of arrival direction distribution. We place ourselves in
the limit of large $N$, therefore replacing the sums over $\omega_i$
by integrals. This leads immediately to

\begin{equation}
\left< C_{\ell} \right> = \frac{1}{4 \pi N_{\mbox{\tiny{eff}}}}\,,
\;\;\;\; \mbox{where}\;\;\;\; N_{\mbox{\tiny{eff}}} =
\frac{N}{1+\left(\frac{\sigma}{\mu}\right)^2}\,.\label{c2}
\end{equation}
}}
\end{itemize}

This last formula shows that if the weight distribution is broad,
i.e. $\sigma/\mu\gtrsim1$,
the effective number of arrival directions becomes small as the only
events which will be ``counted'' are the ones with the largest weights.
This results in an increase of the bias in the power spectrum
estimate in the isotropic case, and the sensitivity to
possible small anisotropies in the model is reduced.

\subsubsection{Effect of Finite-Sized Observer and Simulation Box}

The ideal observer in the method developed here should be point-like,
but, once again for CPU reasons, we modeled the observer as a sphere of
radius $1.5$ Mpc. There is in general no problem with this procedure,
as we expect the UHECR properties such as density, spectrum, composition,
and anisotropies to be roughly the same within this volume. This is
because the sources are extragalactic and typically located at
distances of tens to hundreds of Mpc, and the
magnetic field around the observer is very weak,
$\sim$10 pG. However, in some realizations for the source locations,
one or a few sources can be located within a few Mpc from the observer,
and the finite size of the observer ``spreads'' the image of such
sources over a few degrees. This smoothes out the auto-correlation
function in the first few bins of small angular off-set. For the
case of propagation in a
regime with considerable deflection, this effect is, however, of little
importance.

The finite size of the simulation box can also lead to spurious effects
observable on simulated sky maps. For events at the lowest energies,
$E\sim10\,$EeV, when particles travel over large distances and in the
absence of deflections by magnetic fields, an
excess of events is observed in the direction of the simulation box
corners because of the rectangular geometry of the box, and the
periodic repetition of sources can be observed in the sky maps of
arrival directions within a given source realization.
However, for the cases of interest in our simulations, energies
$E\gtrsim40\,$EeV and/or non-negligible magnetic fields,
these effects disappear completely.

\subsubsection{Specificities of Heavy Nuclei Propagation}

There are a number of difficulties which arise when studying heavy
nuclei and their secondaries in the framework of our method, in
particular concerning the analysis of angular distributions for
simulated events.

\begin{figure}[ht]
\includegraphics[width=0.48\textwidth,clip=true]
{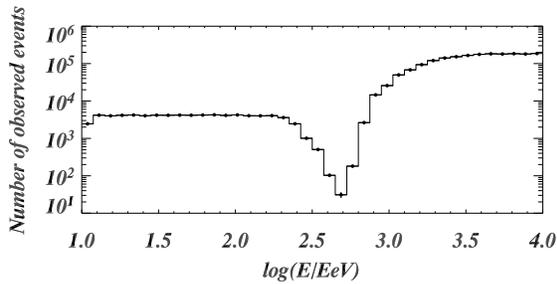}
\caption{Histogram of \emph{injection} energies of events
\emph{detected} by the observer in a scenario with iron injection.
The injection spectrum is flat in the
logarithm from 10 EeV to 10 ZeV, but evidently this is not
the case for the detected events. This effect is explained in the text below.}
\label{fig:Estart_histo}
\end{figure}

\begin{figure}[ht]
\includegraphics[width=0.48\textwidth,clip=true]
{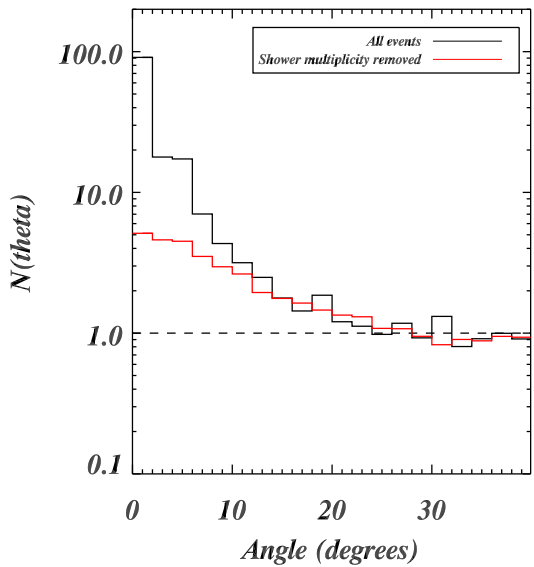}
\caption{Auto-correlation of events with energies above 80 EeV for a
realization of iron
injection with magnetic fields, without (black) and with (red)
correction for nuclear showers arising from the disintegration of a
single iron near the observer. ${\cal N}(\theta)=1$ (dashed line)
corresponds to an isotropic distribution.}
\label{fig:ctheta_showers}
\end{figure}

\begin{itemize}

\item{Propagation of nuclei takes more CPU time than
protons because the deflections are
larger, which leads the Bulirsch-Stoer integrator
to choose smaller stepsizes for integrating the trajectories.}

\item{For heavy nuclei injection, a consequence of photo-disintegration
processes is that particles with a given \emph{observed} energy have
an extremely large range of \emph{injection}
energies. Fig.~\ref{fig:Estart_histo} shows the distribution of
injection energies for recorded events. Two populations appear
clearly: 1) At low injection energies, the flat distribution of iron
originating from nearby sources which is the same as in the case of
proton injection; 2) Nuclei emitted at the highest energies which are
detected as light nuclei after having undergone
photo-disintegration. Due to the approximate conservation of the
Lorentz factor in interactions, this population appears at injection
energies $E\geq A \times E_{min}$, which corresponds to
$\log(E/EeV) \geq 2.75$, as seen in Fig.~\ref{fig:Estart_histo}.
Note that the dip between these two regions is an artifact due to the fact that
iron at those energies undergo frequent photo-disintegrations
but their secondary protons have energies below $E_{\rm min}$
and, therefore, are not detected. This affects
the reconstruction of the injection spectrum in that region. 

As events are reweighted according to their injection energies, the
situation depicted above leads to a broad weight distribution. Therefore, the problems raised
previously are particularly acute for iron injection.
}

\item{Although they are necessary to evaluate the spectrum and
composition, ``showers'' of nuclei which are generated from a single
iron near the
observer create fake signals in the auto-correlation. A crude but
efficient approach to this problem is to use only the first particle
of each shower (composed of 56 particles at most) in the anisotropy
analysis, or equivalently to assign to each particle a weight inversely
proportional to the size of the shower it belongs to. An example
is shown in Fig.~\ref{fig:ctheta_showers}, where the auto-correlation is
computed with and without reweighting of showers, using the
estimator~\cite{lss-protons}
\begin{equation}
{\cal N}(\theta)=\frac{C}{S(\theta)}\sum_{j \neq i}
\left\{\begin{array}{ll}
\omega_i\omega_j & \mbox {if $\theta_{ij}$ in the bin of $\theta$}\\
0 & \mbox{otherwise}
\end{array}\right\}\,.
\label{auto}
\end{equation}
Here, $S(\theta)$ is the solid angle size of the corresponding bin,
and $\omega_i$ are weights as before. The normalization factor
$C=\Omega_e/\sum_{j \neq i}\omega_i\omega_j$,
with $\Omega_e$ denoting the solid angle of the sky region where the
experiment has non-vanishing exposure, is chosen such that an
isotropic distribution corresponds to $N(\theta)=1$.
}

\end{itemize}

We point out that in terms of CPU time consumption, the scenarios explored
in this paper are perhaps the most challenging for obtaining sufficient
statistics. This is the result of combining heavy nuclei propagation
with considerable magnetic fields \emph{and} a discrete distribution
of rare sources. Indeed, these kinds of scenarios have not
been studied yet in great detail because they are computationally
time-consuming. However, it is important to perform simulations and to
predict observables for those scenarios as they allow to scan a large
range of UHECR models.

\begin{table}
\centerline{
\begin{tabular}{|c|c|c|c|c|}\hline
Injected & & & & Number of \\
particles &  $E_{\rm min}$ & $E_{\rm max}$ & EGMF & realizations \\
\hline \hline
Proton & 10 EeV & 1 ZeV & No & 39 \\ \hline % far666_0
Proton & 40 EeV & 1 ZeV & No & 40 \\ \hline % far6661_0
Proton & 10 EeV & 1 ZeV & Yes & 19 \\ \hline % far666
Proton & 40 EeV & 1 ZeV & Yes & 26 \\ \hline %far6661
Iron & 10 EeV & 4 ZeV & No & 100 \\ \hline % ironb_0
Iron & 10 EeV & 4 ZeV & Yes & 56 \\ \hline % ironb
Iron & 10 EeV & 10 ZeV & No & 200 \\ \hline % ironb1_0
Iron & 40 EeV & 10 ZeV & No & 171 \\ \hline % ironb2_0
Iron & 10 EeV & 10 ZeV & Yes & 78 \\ \hline % ironb1
Iron & 40 EeV & 10 ZeV & Yes & 97 \\ \hline % ironb2
\end{tabular}}
\caption{List of simulations carried out. Each simulation corresponds
to $10^4$ recorded ``events''. To increase the statistics at high
energies, some scenarios were simulated twice: once with $E_{\rm min}
= 10$ EeV, and once with $E_{\rm min} = 40$ EeV.}
\end{table}

%%%%%%%%%%%%%%%%%%%%%%%%%%%%%%%%%%%%%%%%%%%%%%%%%%%%%%%%%%%%%%%%%%

\section{Warming up: Proton Injection}

To have a reference and compare with iron simulations, we
extensively simulated proton emission and propagation with parameters
similar to the ``most-favored'' scenario number 6 of
Ref.~\cite{lss-protons}. In the framework of this scenario, for a
source density of $2.4\times 10^{-5} \mbox{Mpc}^{-3}$, we
chose a distribution of sources
following the baryon density, with and without EGMF. The maximum
injection energy was set to 1 ZeV$=10^{21}\,$eV, and all events
above $E_{\rm min}=10\,$EeV are taken into account. The injection spectrum in
this energy range was chosen $\propto E^{-\alpha}$, where the average
value of $\alpha$ is fitted to the data and typically varies
between $\simeq2$ and $\simeq2.4$.

The main conclusions from these simulations are:

\begin{figure}[ht]
\includegraphics[width=0.48\textwidth,clip=true]{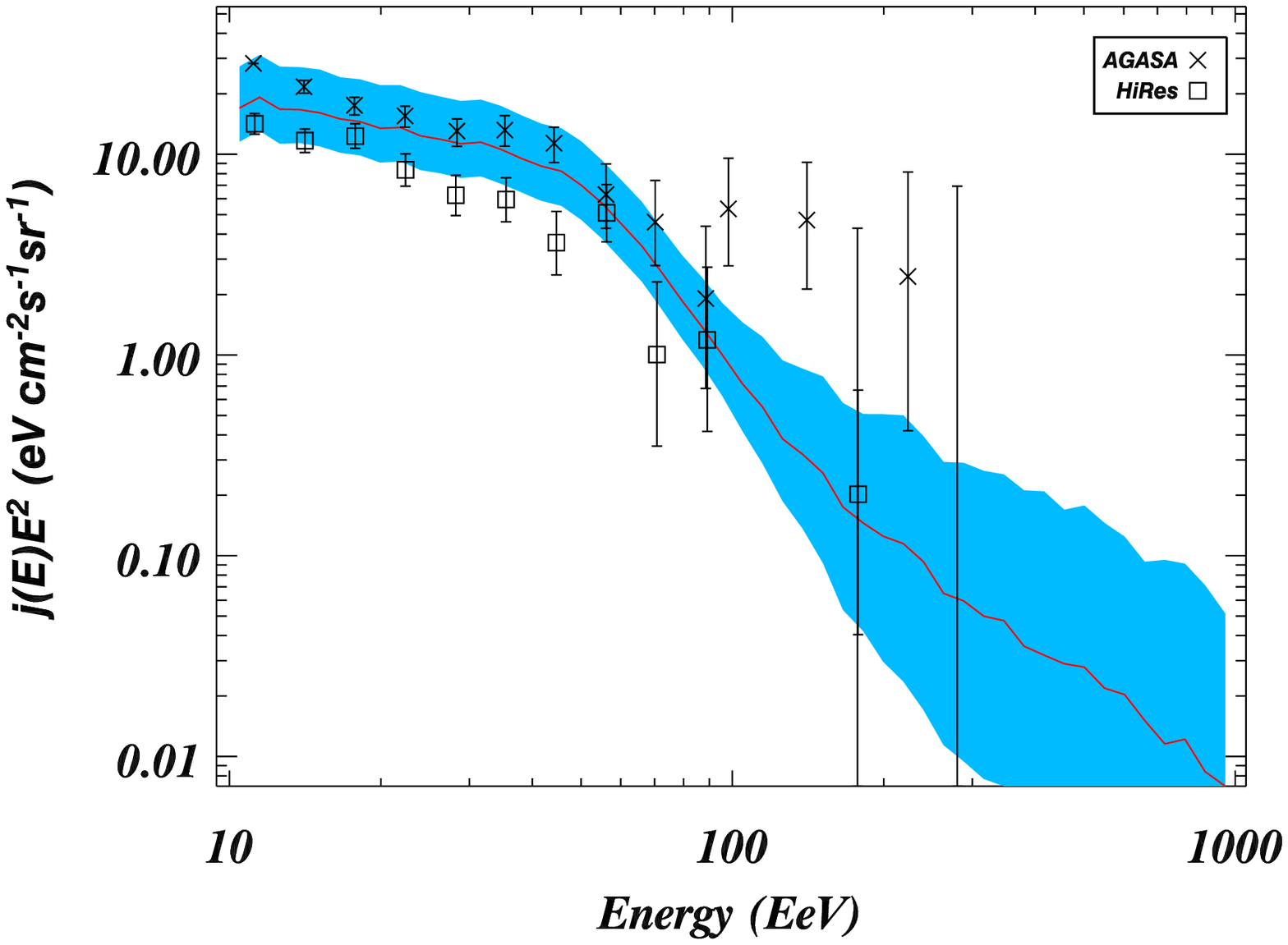}
\includegraphics[width=0.48\textwidth,clip=true]{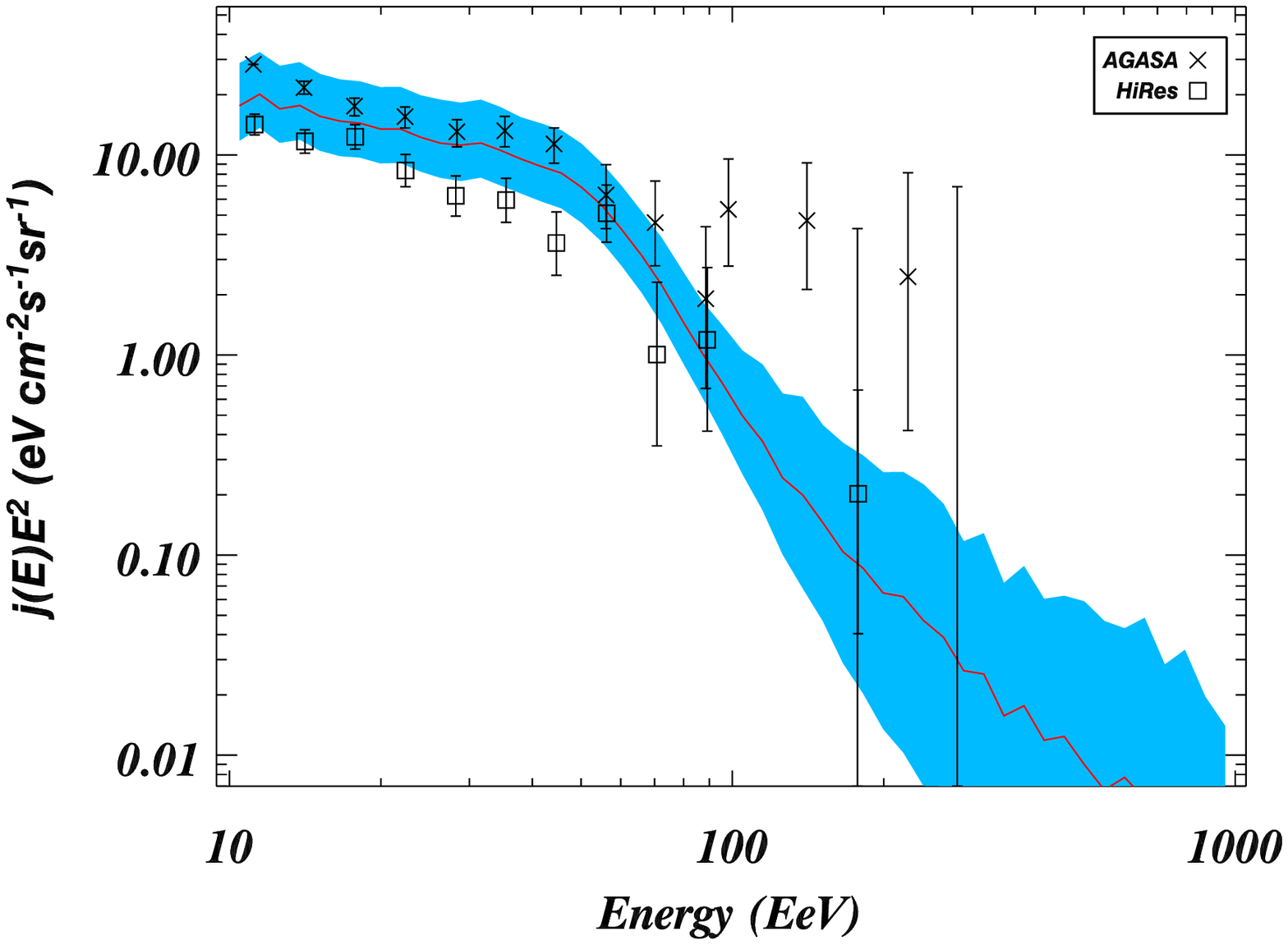}
\caption{Energy spectra predicted in the case of proton injection at
the sources, without
(top panel) and with (bottom panel) EGMF. The average (solid line) and
cosmic variance (shaded band) result from various realizations of
the source properties and locations. The source positions are
assumed to have a statistical distribution proportional to the baryon density.
For each source location, 50 realizations for
source intensity and spectral index are drawn according to
the distributions of Eq.~(\ref{source_fluct}). The mean source spectral
index is $\langle\alpha\rangle = 2.4$. To guide the eye
we also show the spectra measured by AGASA~\cite{agasa-spectrum}
and HiRes~\cite{hires-spectrum}. The normalization of our curves
is obtained by adjusting the average spectrum to low energy data.}
\label{fig:pspectrum_bfield}
\end{figure}

\begin{figure}[ht]
\includegraphics[width=0.48\textwidth,clip=true]
{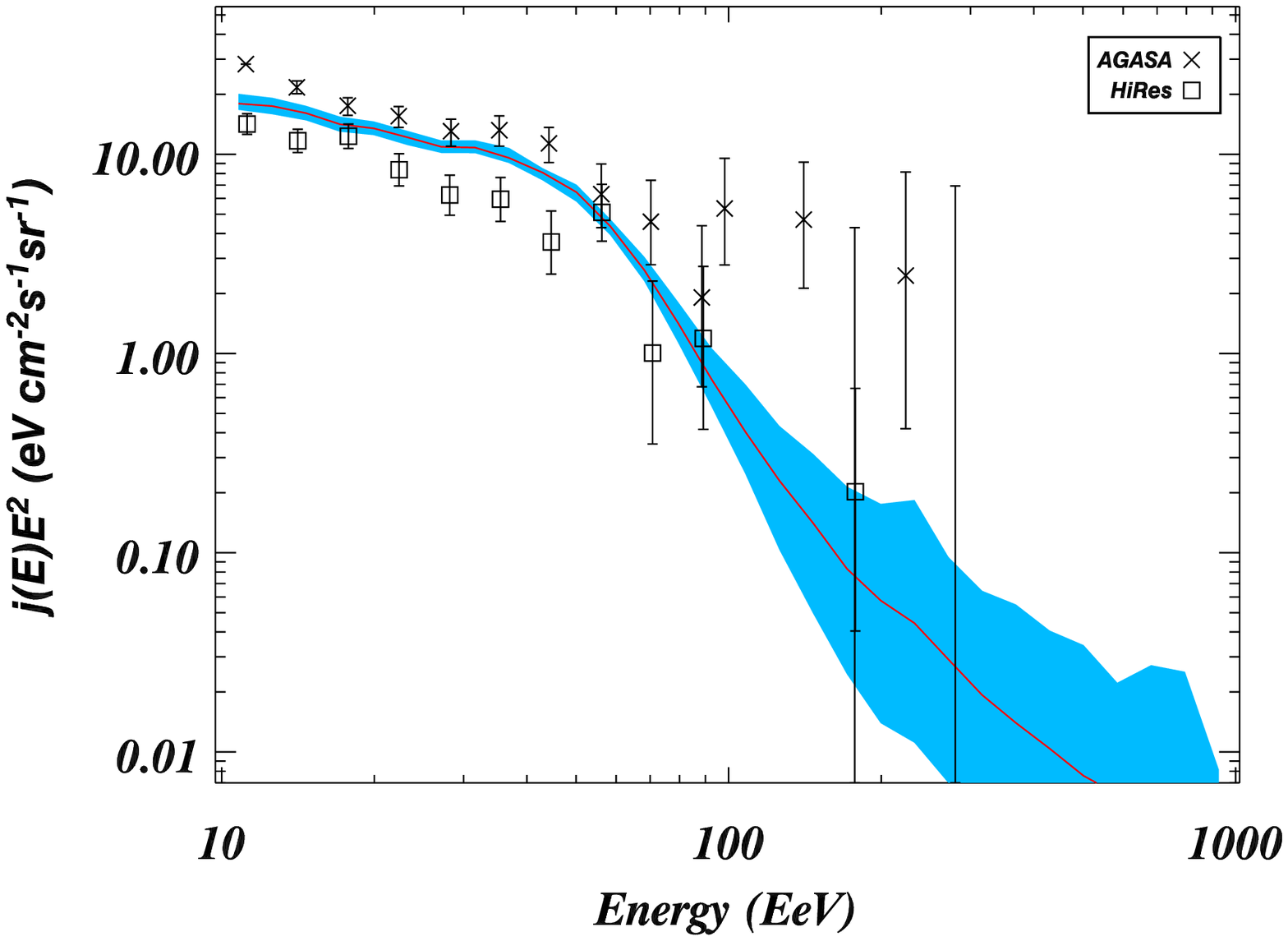}
\includegraphics[width=0.48\textwidth,clip=true]{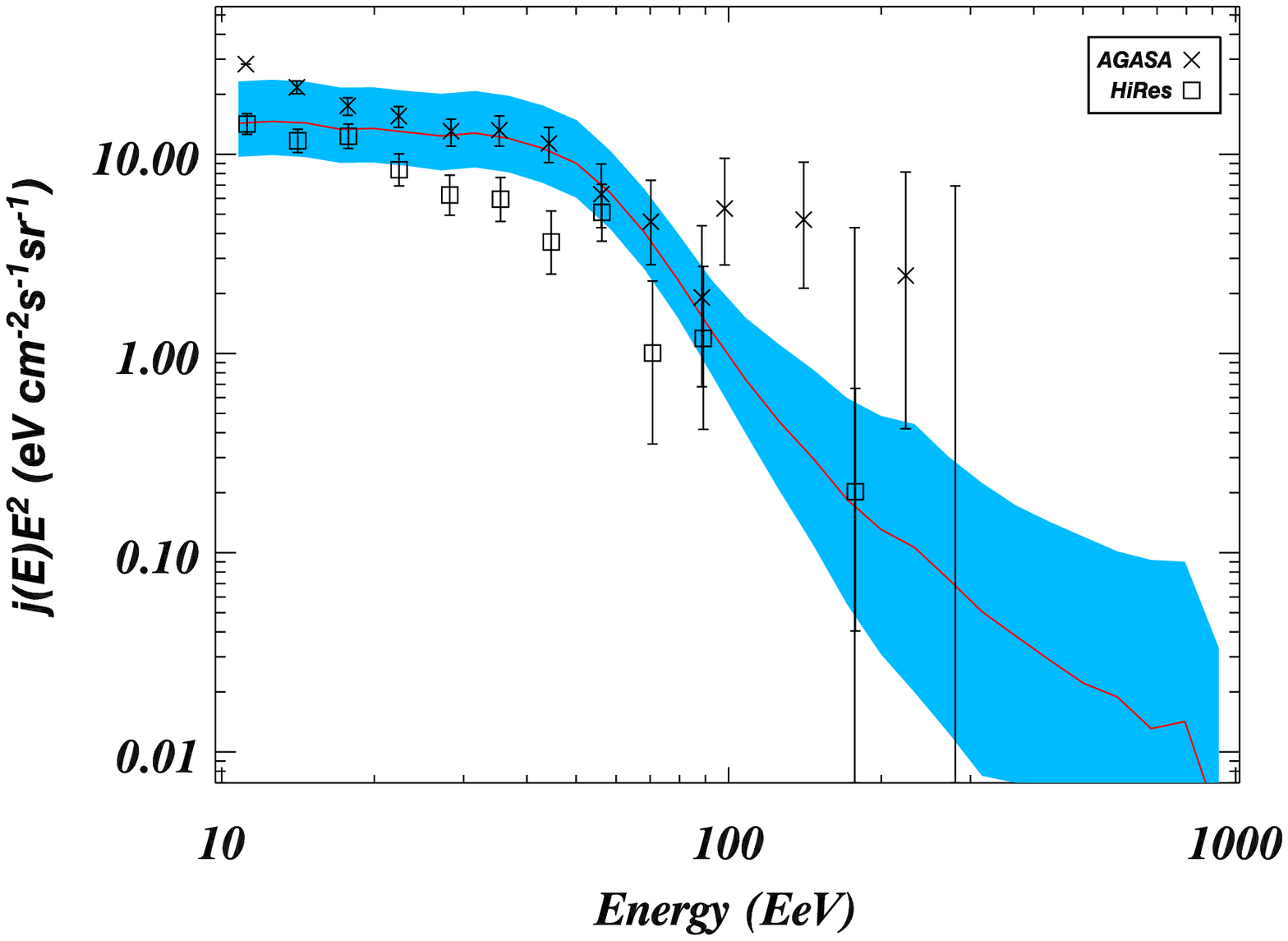}
\caption{Same as Fig.~\ref{fig:pspectrum_bfield}, for proton
injection with EGMF fields: In the top panel all sources have the same
intensity $Q = 1$ and spectral index $\alpha = 2.4$, i.e. cosmic
variance is only due to fluctuations in the source locations.
The bottom panel includes source property fluctuations as in
Fig.~\ref{fig:pspectrum_bfield}, but with a mean injection spectral index
$\langle \alpha \rangle = 2.0$ instead of 2.4.}
\label{fig:pspectrum_fluct}
\end{figure}

\begin{figure}[ht]
\includegraphics[width=0.48\textwidth,clip=true]{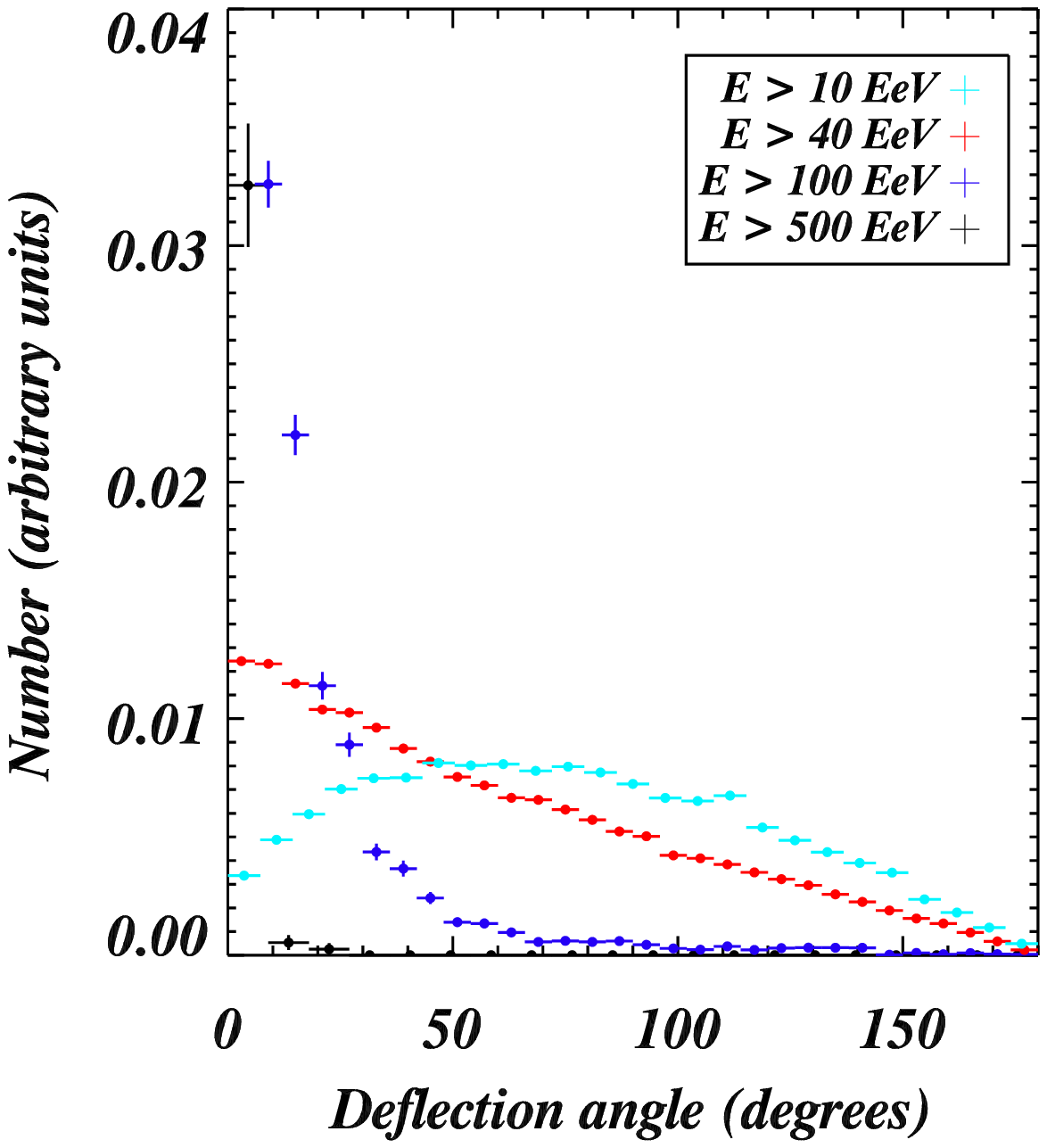}
\includegraphics[width=0.48\textwidth,clip=true]{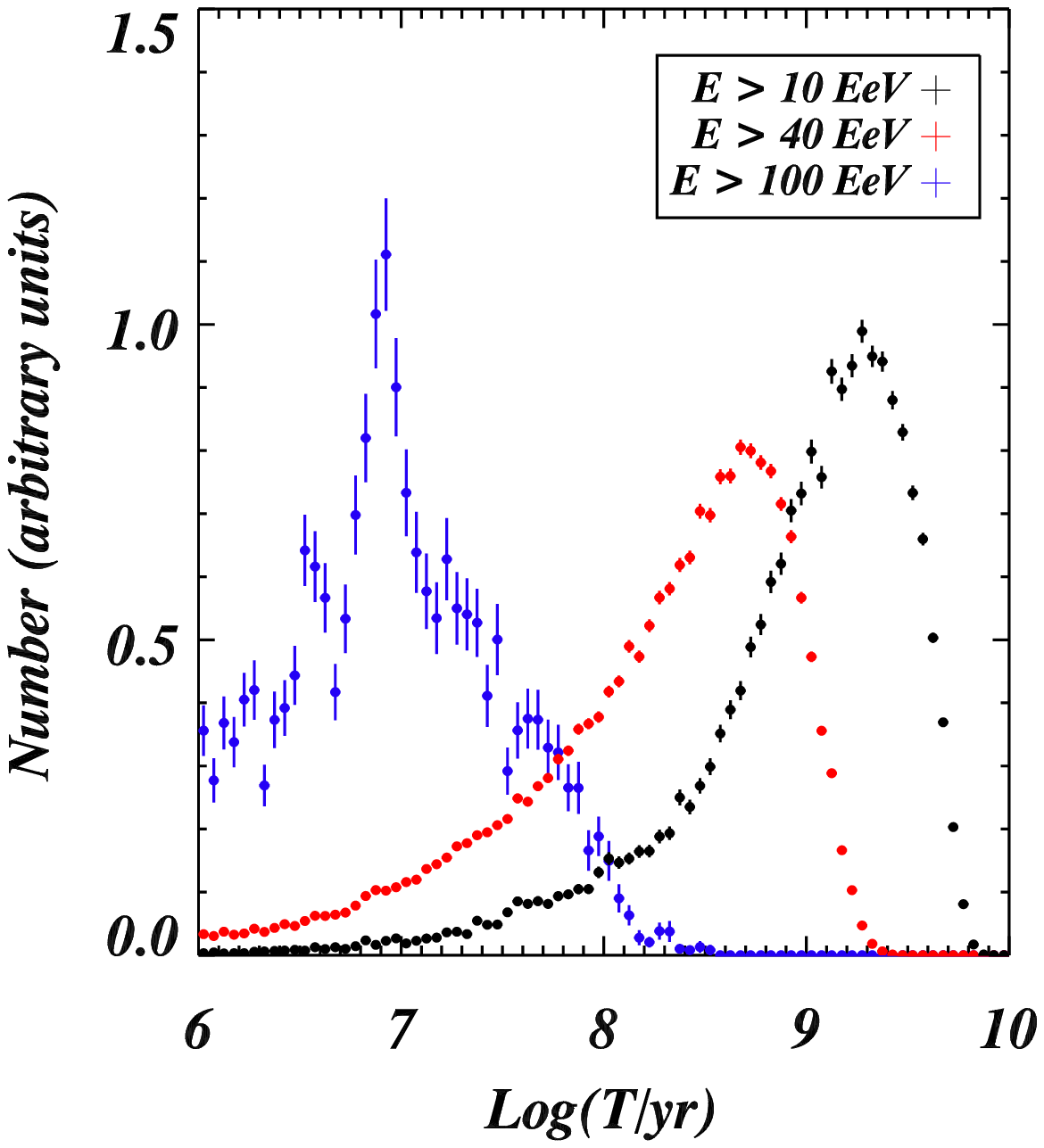}
\caption{Distributions of deflection angles (top panel) and time delays
versus straight-line propagation time (bottom panel) of simulated
trajectories for proton injection with spectral index $\alpha=2.4$.
The distributions are cumulated over all source location
realizations and the error bars are drawn from Poissonian
statistics and reflect the finite number of simulated trajectories.}
\label{fig:p_deflec}
\end{figure}

\begin{figure}[ht]
\includegraphics[width=0.48\textwidth,clip=true]{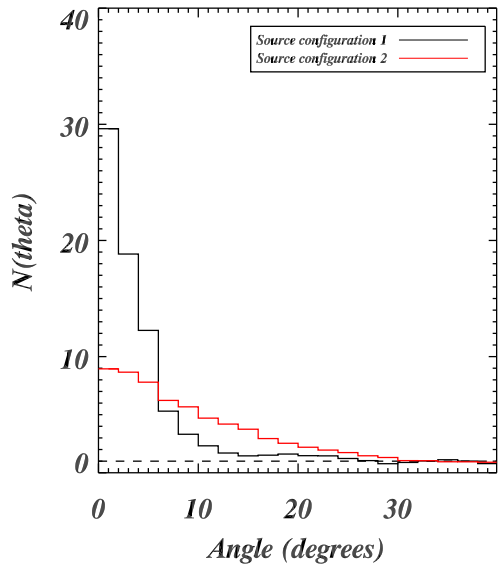}
\caption{Examples of auto-correlation for $E > 100$ EeV for proton
injection with spectral index $\alpha=2.4$ and EGMF, for two
different source configurations. Within the same model,
fluctuations of the distances to the nearest sources from the observer
generate extremely different auto-correlation (and therefore
clustering) shapes. $N(\theta)=1$ (dashed line) corresponds to an
isotropic distribution.}
\label{fig:p_example_autocorr}
\end{figure}

\begin{itemize}

\item{\textbf{The predicted energy spectra} are shown in
Figs.~\ref{fig:pspectrum_bfield} and~\ref{fig:pspectrum_fluct}.
As expected, in this astrophysical scenario the GZK feature appears
at energies above $\sim4\times 10^{19}\,$eV. Since the sources are distributed
at various distances from the observer, the feature does not appear as
a sharp cut-off, but rather as a steepening of the spectral index. Let us
remark that we are not able to study the ankle within these simulations
which are limited so far to energies above 10 EeV. A proton average
injection spectral index $\langle\alpha\rangle = 2.4$
fits better than $\langle\alpha\rangle = 2.0$ to current AGASA/HiRes
data at $10 < E < 40\,$EeV, see Fig.~\ref{fig:pspectrum_fluct}.

A major aspect of the obtained spectra is
the presence of a large cosmic variance, which has different origins
below and above the GZK feature: Below the GZK feature, this variance
is mostly due to source luminosity fluctuations, since it almost
disappears if sources have identical properties, see
Fig.~\ref{fig:pspectrum_fluct}. Above the GZK feature, fluctuations of
the source locations also significantly contribute to cosmic variance: The
spectrum at post-GZK energies depends substantially on the presence
of a few powerful sources close to the observer. This is because our source
density is low: No continuous approximation for the source distribution
could reproduce such results.

Finally, the comparison of the upper and lower panels in
Fig.~\ref{fig:pspectrum_bfield} demonstrates the effect of EGMF on
the spectrum: The attenuation of spectra at high energies is
slightly more pronounced
in the case of EGMF, the mean spectral slope in the energy range
between~$\simeq50\,$EeV and~$\simeq200\,$EeV being~$\sim 6.4$
instead of~$\sim5.3$. This is due to the increase of the average
traveled distance in the presence of EGMF.}

\item{\textbf{Deflections and time delays} are defined here as the
deflection and delay accumulated by UHECRs compared to photons.
In general they are very difficult to observe, but $i$) moderate deflections
from an unambiguously identified source could be measured, and $ii$) time
delays could be measured in case a GRB or a SN explosion can be
identified as source. Predicted deflection angle and time delay
histograms are shown in Fig.~\ref{fig:p_deflec}. It can be seen that
above 10 EeV the deflection distribution obtained is not trivial, i.e.
it is not the one we would obtain if particles were in the fully
diffusive regime. One can also remark that even at 100 EeV, typical
deflections are still of the order of 10-40 degrees. This is due to
the fact that EGMF are substantial, and sources are generally located
in magnetized areas: particles are mostly deflected in the local
environment around their sources.
The time delay histogram shows that typical time
delays are large, around 1 Gy at 10 EeV, due to the EGMF.
Thus in this scenario, even at the highest energies time
delays will be too large to be directly measurable.}

\item{\textbf{Auto-correlation of events.} Even for substantial
EGMF, the fact that at high energies all events come from a few sources
leads to strong auto-correlation
signals. A major feature is, however, the high cosmic variance of this
auto-correlation, which depends strongly on the positions and luminosities
of the sources closest to the observer. This can be seen in
Fig.~\ref{fig:p_example_autocorr} which shows two different source
realizations: The first one presents a smooth
auto-correlation extending to 30 degrees, whereas in the second
realization, with a nearby
source, the auto-correlation is highly peaked, with a gaussian-like
shape of width $\sigma \sim 5$ degrees.}

\end{itemize}

%%%%%%%%%%%%%%%%%%%%%%%%%%%%%%%%%%%%%%%%%%%%%%%%%%%%%%%%%%%%%

\section{Iron Injection Without Magnetic Deflection}

This section illustrates general features of heavy nuclei
propagation. We do not take into account
particle deflections, and, therefore, these results can be
compared to the case of small deflections obtained in the
EGMF scenarios of Refs.~\cite{dolag,yamamoto}. Note, however, that
deflection can still be of the order of $\sim20^\circ$
in the EGMF scenario of Ref.~\cite{dolag} when their results are
extrapolated to nuclei.

\subsection{Spectrum}

\begin{figure}[ht]
\includegraphics[width=0.48\textwidth,clip=true]{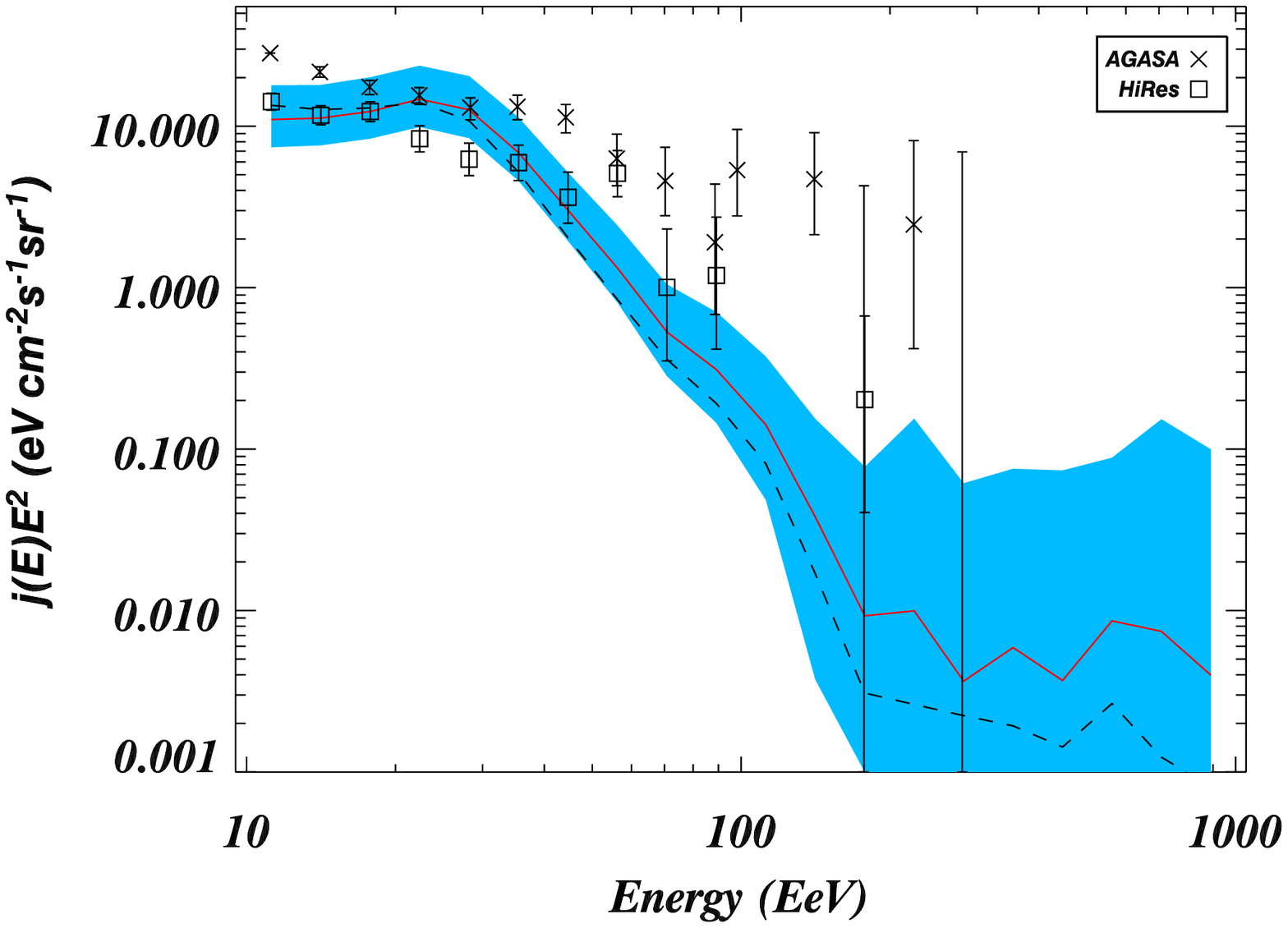}
\includegraphics[width=0.48\textwidth,clip=true]{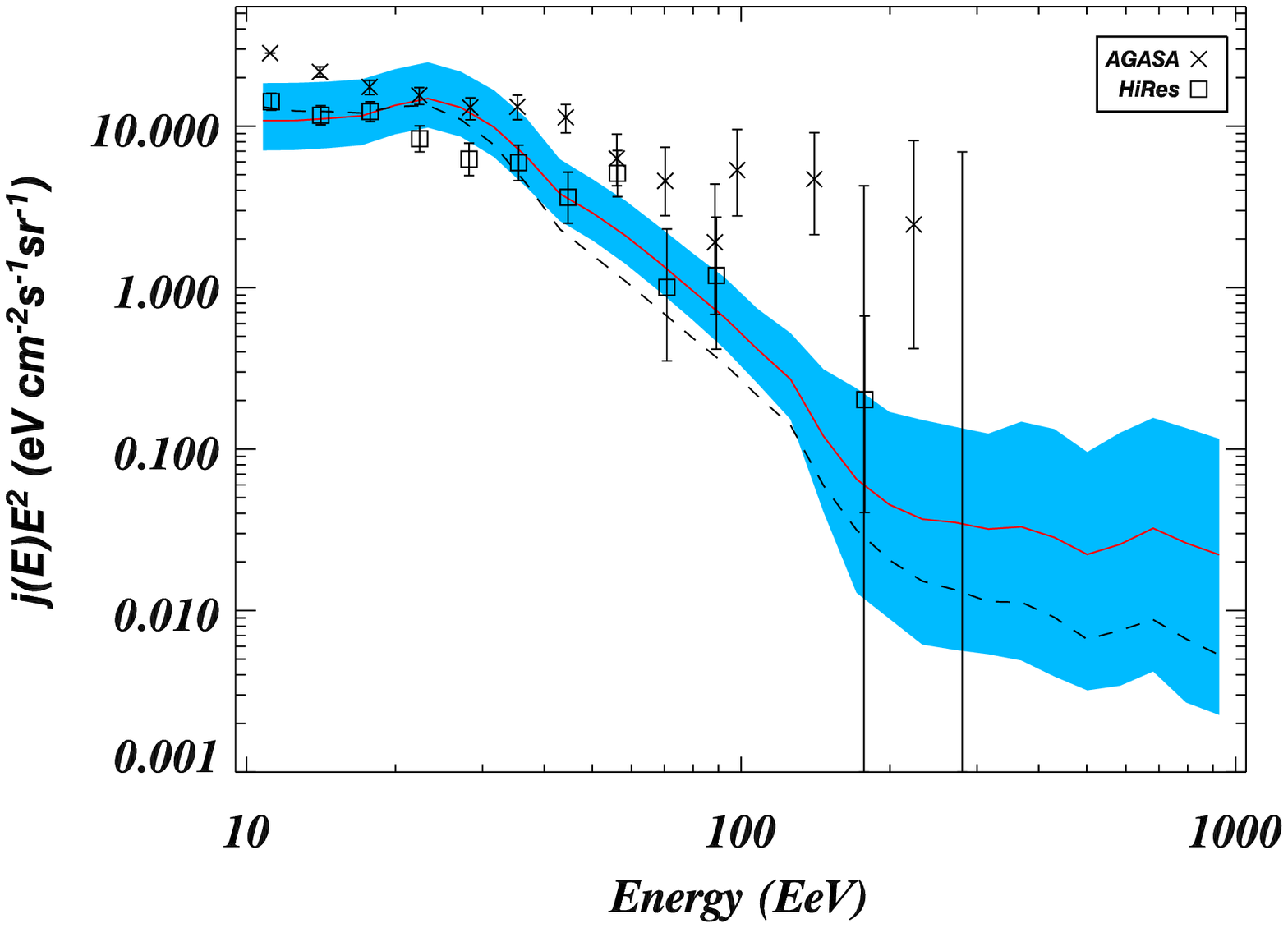}
\caption{Energy spectra predicted for iron injection without deflection.
The average (solid line) and cosmic variance (shaded band) are
obtained as in Fig.~\ref{fig:pspectrum_bfield}.
The average source spectral index is $\langle\alpha\rangle = 2.0$,
whereas the dashed lines show the average spectrum for
$\langle\alpha\rangle = 2.4$.
The maximum injection energy is set to 4 ZeV (top panel) and 10 ZeV
(bottom panel). }
\label{fig:iron_spectrum_nob}
\end{figure}

In Fig.~\ref{fig:iron_spectrum_nob} we represent the spectra
obtained with an average iron injection spectral index
$\langle\alpha\rangle=2$, and two different
maximum injection energies. The injection spectra which
allow to fit the observed spectra tend to be somewhat
harder than in the case of protons. Clearly, however, the spectral
shape predicted at energies $E\lesssim30\,$EeV is not compatible with
observations. At those energies, a lighter component, e.g. from
proton injection, is necessary. Around 20-30 EeV a small ``bump''
is predicted for pure iron injection which is mostly due to secondary
sub-GZK protons.

At energies $E\gtrsim30\,$EeV we observe a flux suppression due to nuclear
photo-disintegration. The shape of the cut-off is not the same as in
Fig.~\ref{fig:pspectrum_bfield} for proton injection: in particular
it starts at lower energies and it is flatter. Furthermore, there is a
significant steepening around 100 EeV, with a flattening between $10^{20}$
and $10^{21}$ eV. The location of this flattening or ``spectrum
recovery'' strongly depends on the source configuration, as can be
seen from the large cosmic variances in
Fig.~\ref{fig:iron_spectrum_nob}, and also on the maximal injection
energy: Fig.~\ref{fig:iron_spectrum_nob} shows that the recovery flux
is statistically higher when increasing $E_{\rm max}$.

%%%%%%%%%%%%%%%%%%%%%%%%%%%%%%%%%%%%%%%%%%%%%%%%%%%%%%%%%%%%%%%%%%%%%%%%%%

\subsection{Composition}

\begin{figure}[ht]
\includegraphics[width=0.48\textwidth,clip=true]{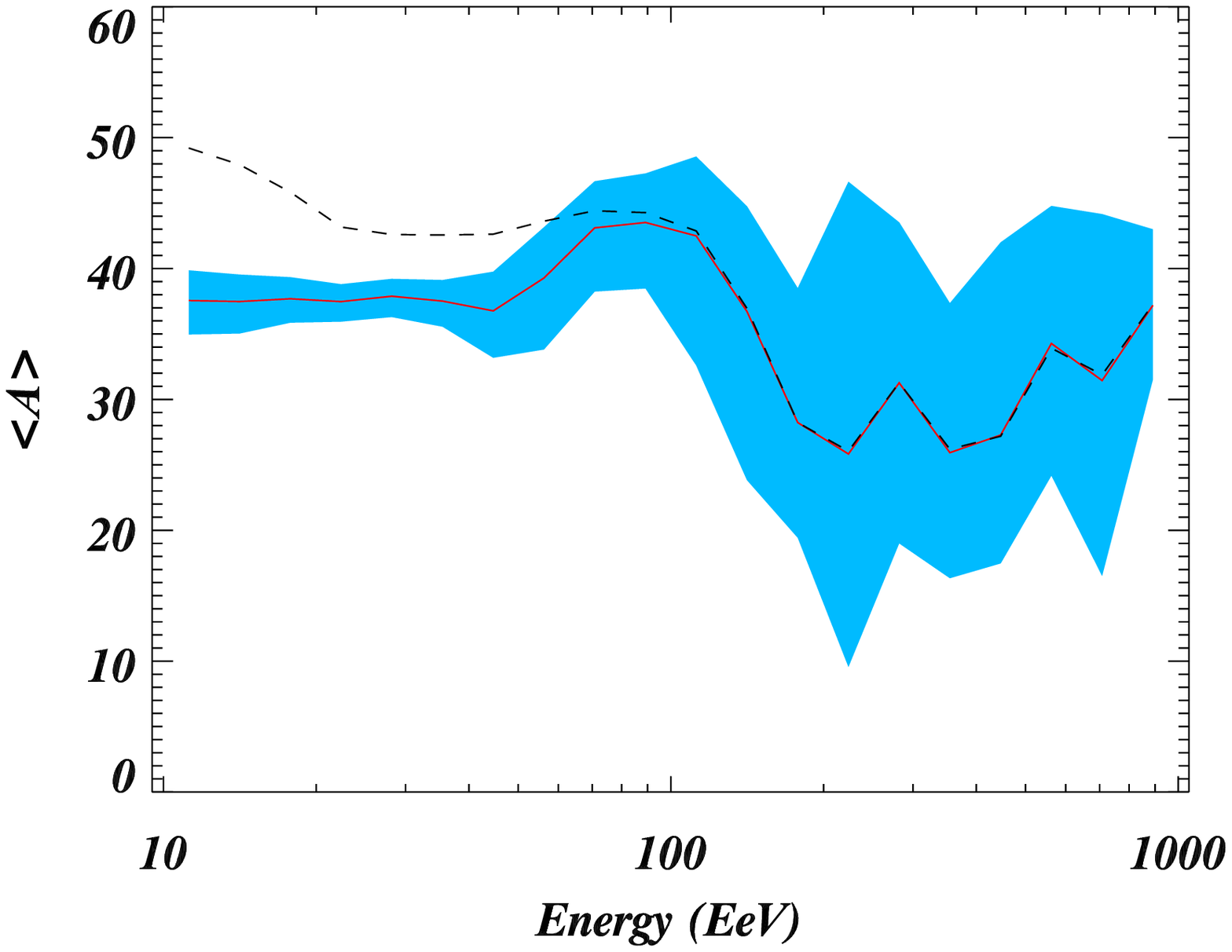}
\includegraphics[width=0.48\textwidth,clip=true]{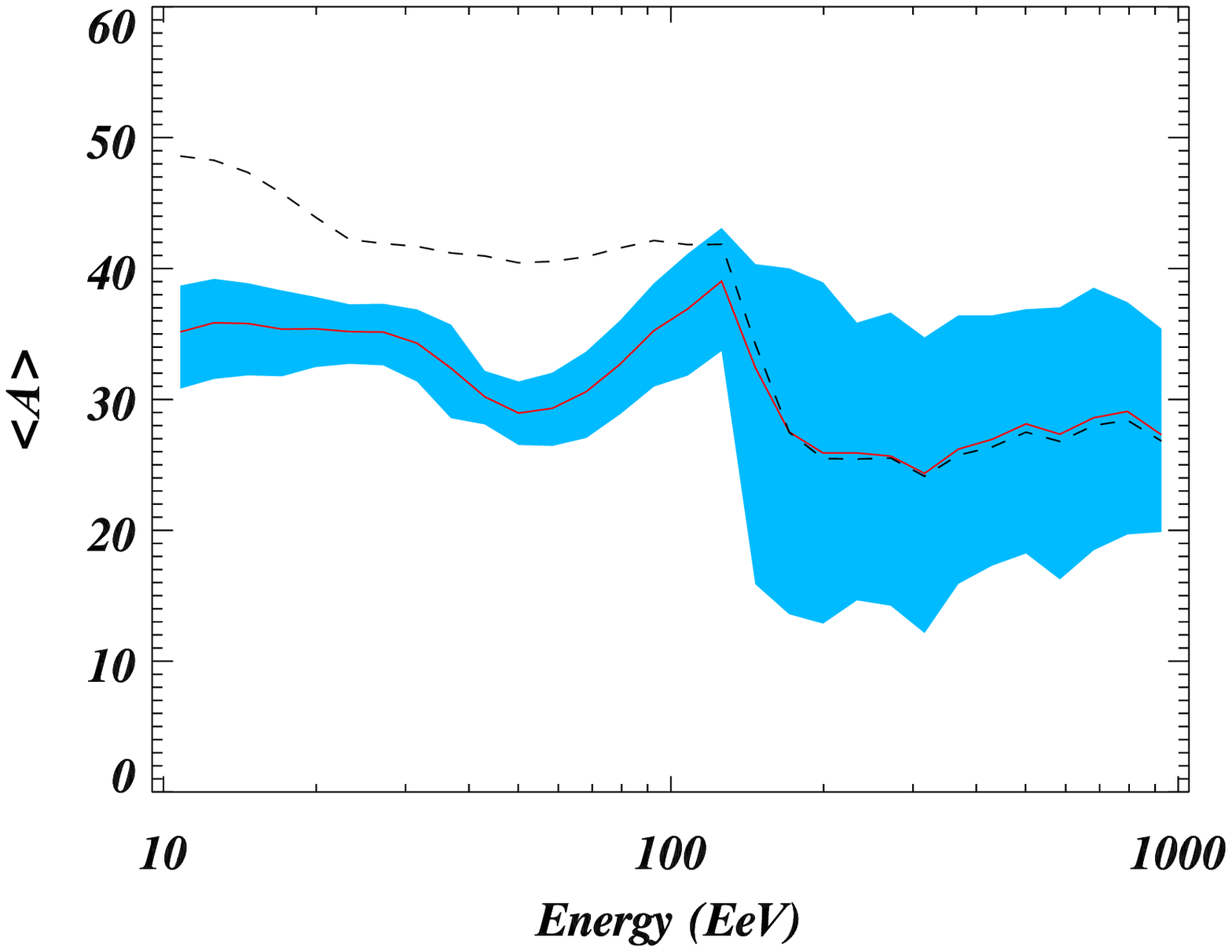}
\caption{Predicted average atomic mass $\langle A\rangle$ as a function
of observed particle energy corresponding to the spectra shown
in Fig.~\ref{fig:iron_spectrum_nob} for iron injection without deflection.
Mean and cosmic variance are computed as in Fig.~\ref{fig:iron_spectrum_nob}
with $\langle\alpha\rangle = 2.0$ and maximum injection energy of
4 ZeV (top panel) and 10 ZeV (bottom panel). The dashed lines show the
average composition for $\langle\alpha\rangle = 2.4$.}
\label{fig:iron_compo_nob}
\end{figure}

\begin{figure}[ht]
\includegraphics[width=0.48\textwidth,clip=true]{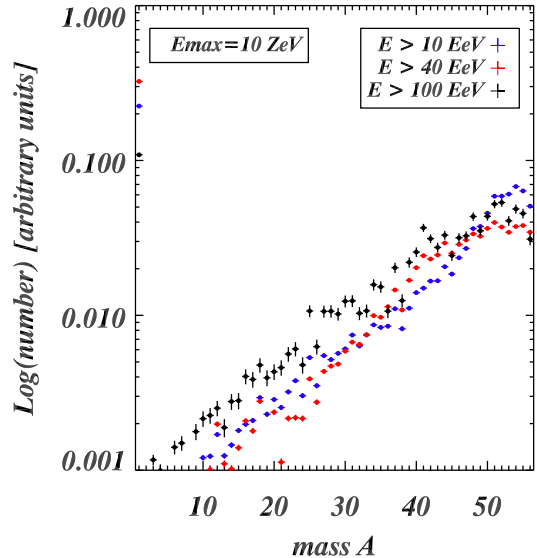}
\caption{Distributions of masses in the case of iron injection with
spectral index $\alpha = 2$, $E_{\rm max} = 10$ ZeV and no
EGMF. The distributions are cumulated over various source position
realizations and the error bars are drawn from Poissonian
statistics reflecting the finite number of simulated trajectories.}
\label{fig:mass_histo}
\end{figure}

The mean composition as a function of energy is shown in
Fig.~\ref{fig:iron_compo_nob}, where the cosmic variance is due to the
fluctuations in source locations and properties.
At energies $E\lesssim30\,$EeV, the mean mass $\langle A \rangle$ is
typically higher than 35, reflecting in fact a bimodal distribution: At
these energies we have on the
one hand protons originating from high-energy heavy nuclei
photo-disintegration, and on the other hand heavy particles
injected and surviving propagation at low energies.
If we model the composition as a simple (H,Fe) mixture, which is done
in most experimental studies at these energies, the observed $\langle
A \rangle \sim 35$ in Fig.~\ref{fig:iron_compo_nob} corresponds to
$\simeq 40$\% H and $\simeq 60$\% Fe even for pure iron injection.
The dashed lines in Fig.~\ref{fig:iron_compo_nob} represent  $\langle
A \rangle$ in the case of a different injection spectral index,
$\langle \alpha \rangle = 2.4$ instead of 2: It is evident that
the steeper the injection spectrum, the heavier the
composition. Indeed, decreasing $\alpha$ enhances the number of high
energy injected iron and, therefore, of low-energy proton secondaries
which implies a lighter composition at energies below $\simeq100\,$EeV.

Between $\simeq40\,$EeV and $\simeq200\,$EeV, the situation is less
clear as there is a competition between two phenomena: On the one
hand, at energies above $\sim100\,$EeV iron is efficiently
photo-disintegrated, whereas below $E_{\rm max}/56$ secondary protons
from dissociated iron nuclei
appear. Both effects suppress the average atomic mass $\langle A \rangle$.
This can lead to a bump around $\simeq100\,$ EeV provided that the two
effects operate in different energy ranges, that is if
$E_{\rm max}/56 $  is well below $\sim100\,$EeV, as in the top panel
of Fig.~\ref{fig:iron_compo_nob}.
In general, these effects are reduced for steeper injection
spectral index $\alpha$, as can be seen in Fig.~\ref{fig:iron_compo_nob},
because there are fewer secondary products to affect a larger population of
particles in the lower energy part of the spectrum.
Therefore, at these transition
energies, the mean composition appears to be subject to
large fluctuations depending on various parameters.

Above $\simeq200\,$EeV, the mean value of atomic mass $\langle A\rangle$
depends substantially, like the spectrum, on the source locations.
However, in general the average $\langle A \rangle$ slightly increases
with energy, resulting from the kinematic condition $E_{\rm max}(A) = E_{\rm
max}(56)\times A /56$, as photo-disintegration roughly conserves the
Lorentz factor.

We also found that at all energies considered, fluctuations of
source luminosities and injection spectra make an insignificant
contribution to the cosmic variance in the distribution of
$\langle A \rangle$.

\begin{figure}[!ht]
\includegraphics[width=0.48\textwidth,clip=true]{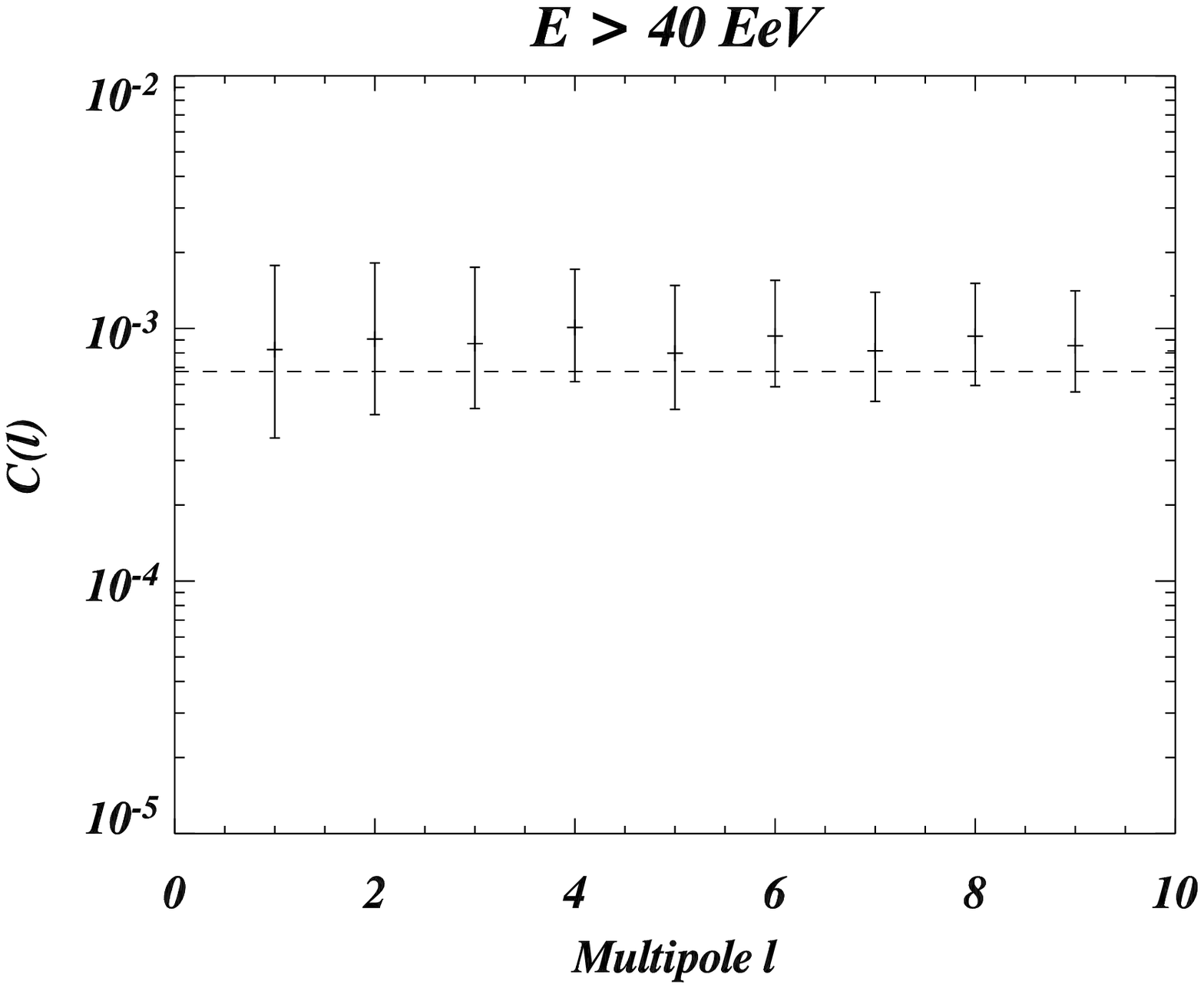}
\includegraphics[width=0.48\textwidth,clip=true]{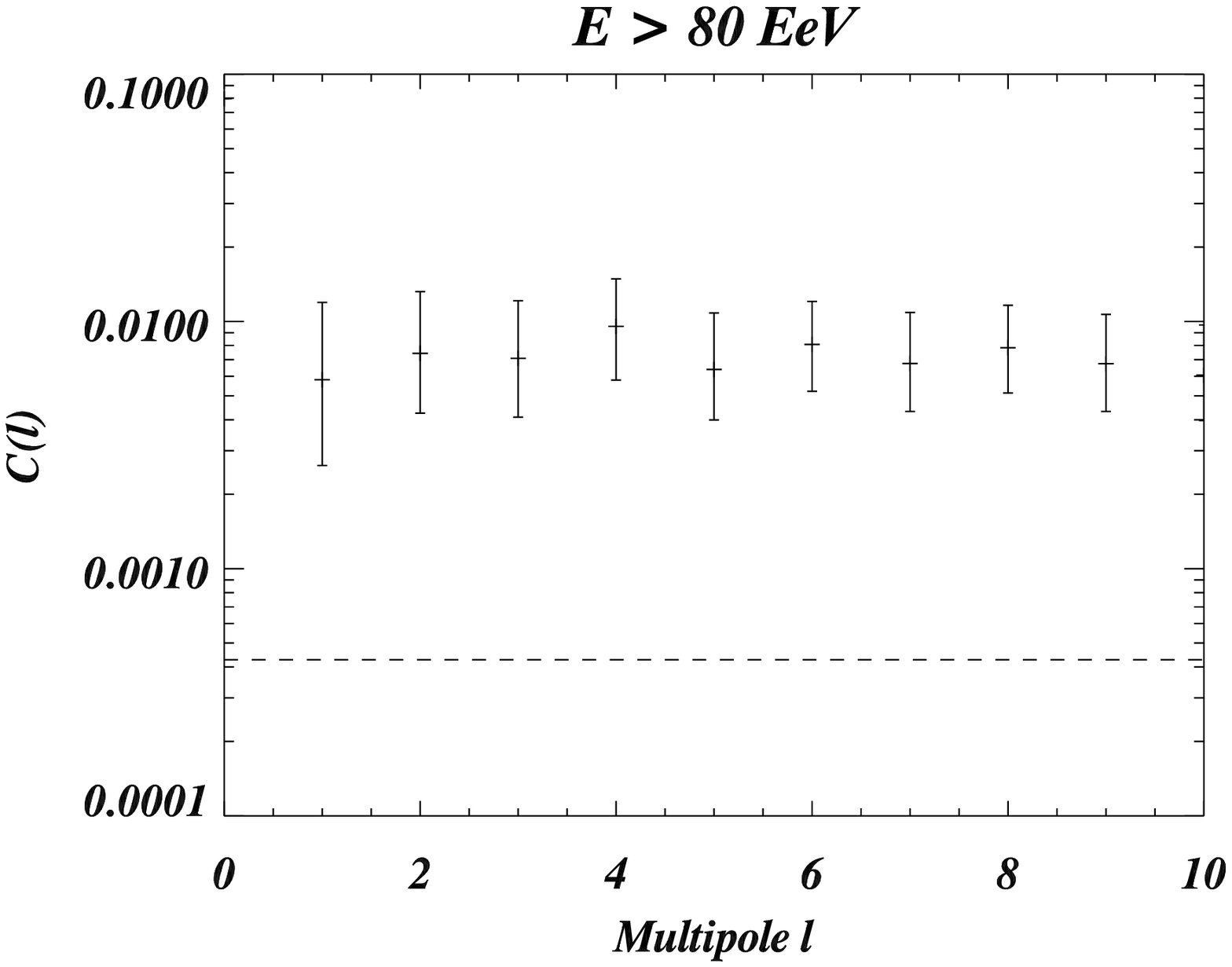}
\caption{Predictions for the angular power spectrum in the case of
iron injection without any magnetic field, above 40 (top) and 80 EeV
(bottom). The dashed line, computed from Eq.~(\ref{c2}), corresponds
to an isotropic distribution of $10^4$ weighted simulated trajectories
above 40 EeV. The error bars represent the cosmic variance
obtained by simulating various source positions and properties.}
\label{fig:iron_cl_nob}
\end{figure}

\begin{figure}[!ht]
\includegraphics[width=0.48\textwidth,clip=true]{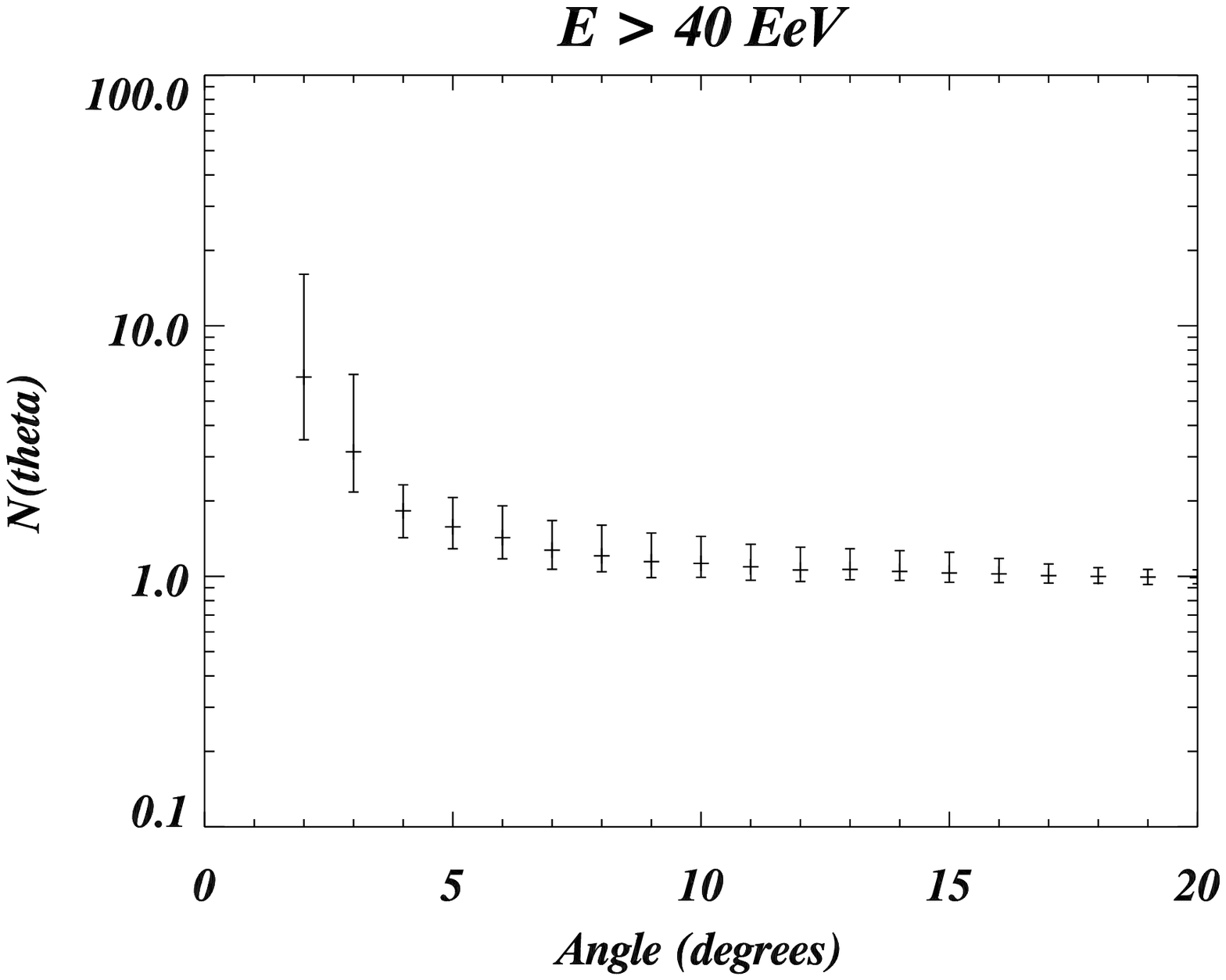}
\includegraphics[width=0.48\textwidth,clip=true]{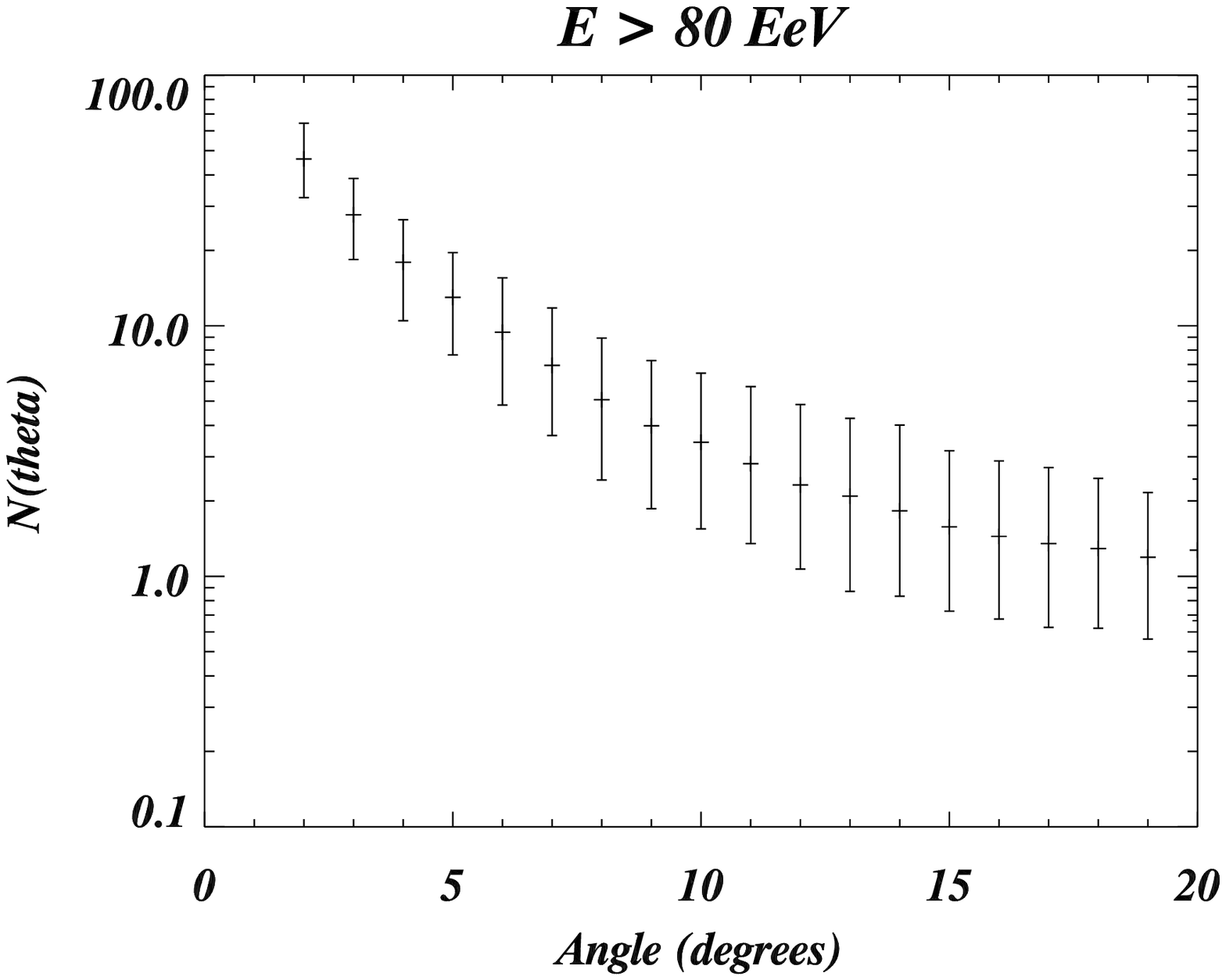}
\caption{Predictions for the auto-correlation function of observed
events in the case of iron injection without EGMF, above 40 and 80
EeV. We do not show the auto-correlation in the first bin and showers
have been corrected for, see Fig.~\ref{fig:ctheta_showers}.
The error bars represent the cosmic variance obtained by
simulating various source positions and properties. ${\cal N}(\theta)
= 1$ corresponds to an isotropic distribution.}
\label{fig:iron_autocor_nob}
\end{figure}

In Fig.~\ref{fig:mass_histo}, we present the distributions of mass A
for pure iron injection, in the case of $\alpha = 2$ and $E_{\rm max}
= 1$ ZeV. There are two populations of detected events, namely
a contribution of protons on the one hand, and a broad
distribution of heavy nuclei dominated by the iron group elements on the other
hand. This bimodal distribution is also reflected in
Fig.~\ref{fig:Estart_histo} and justifies to consider the
distributions as a (H,Fe) mixture at first approximation, as was done above. 
Fig.~\ref{fig:mass_histo} shows a nuclear cascade in which the
abundance of a given nucleus relative to its parent nucleus is
governed by the ratio of the photo-disintegration rate of the parent
and its total disappearance rate. For unstable daughter nuclei, this
ratio is in general smaller than one, which explains the roughly
exponential behavior of this cascade.

We note that the predicted composition discussed in this section is
not ruled out by the still very scarce experimental data~\cite{watson-compo}.

%%%%%%%%%%%%%%%%%%%%%%%%%%%%%%%%%%%%%%%%%%%%%%%%%%%%%%%%%%%%%%%%%%%%%%%%%%

\subsection{Anisotropies}

In Figs.~\ref{fig:iron_cl_nob} and~\ref{fig:iron_autocor_nob}, we
represent the autocorrelation function and angular power spectrum
predicted in the case of iron injection without any magnetic
deflection. As for the case of spectrum and composition, fluctuations
due to various source positions and properties are computed: They are
rather large, but at energies $E\gtrsim40\,$EeV they are smaller
than the deviation from the values predicted from isotropy.

In the null hypothesis of isotropy, the auto-correlation ${\cal
N}(\theta)$ is 1 by definition, and the angular power spectrum
$C_{\ell}$ is
flat, at a value depending on the number of recorded events and on
their weights, computed according to Eq.~(\ref{c2}). The
isotropic level of angular power spectrum is represented by the dashed
lines in the plots.

The angular power spectrum is approximatively flat as is expected for
a distribution of point sources: The sky map being like a sum of Dirac
$\delta$ functions, its Fourier transform is flat.
At 40 EeV, sources contribute up to cosmological distances
and therefore the $C_{\ell}$ are compatible with the
expected value from isotropy. At 80 EeV however, because the number of
observed sources becomes small, and the predicted $C_{\ell}$ is at a
significantly higher level than expected from isotropy.

As expected in the absence of any magnetic field, the auto-correlation
of events is sharply peaked at small angles, reflecting also the small
source density. The auto-correlation signal is larger at 80 EeV than
at 40 EeV, as the number of effectively contributing sources becomes
smaller. However, the peak in the auto-correlation at
small angles has a finite width because $i$) sources are
concentrated in regions of large baryonic density, which makes them
often appear next to each other, and $ii$) there is an artificial
spreading of the auto-correlation over a few degrees due to the
finite size of the observer, see Sect.~II-B. The power spectra and
auto-correlations are comparable to the analogous case with proton
injection~\cite{lss-protons}, as
might be expected for negligible deflection.

%%%%%%%%%%%%%%%%%%%%%%%%%%%%%%%%%%%%%%%%%%%%%%%%%%%%%%%%%%%%%%%%%%%%%%%%%%

\section{Iron Injection With Magnetic Deflection}

\subsection{Spectrum and Composition}

\begin{figure}[ht]
\includegraphics[width=0.48\textwidth,clip=true]{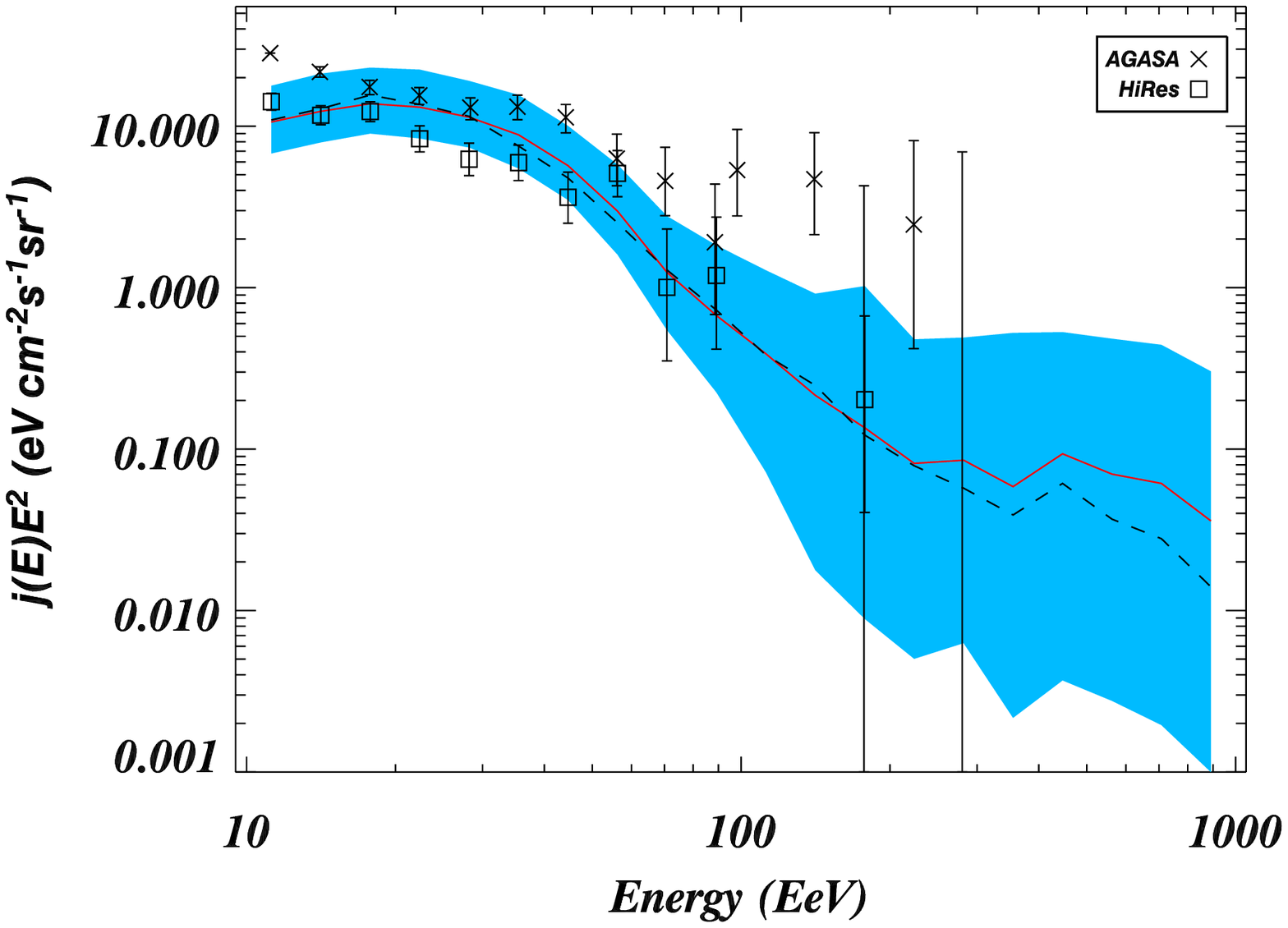}
\includegraphics[width=0.48\textwidth,clip=true]{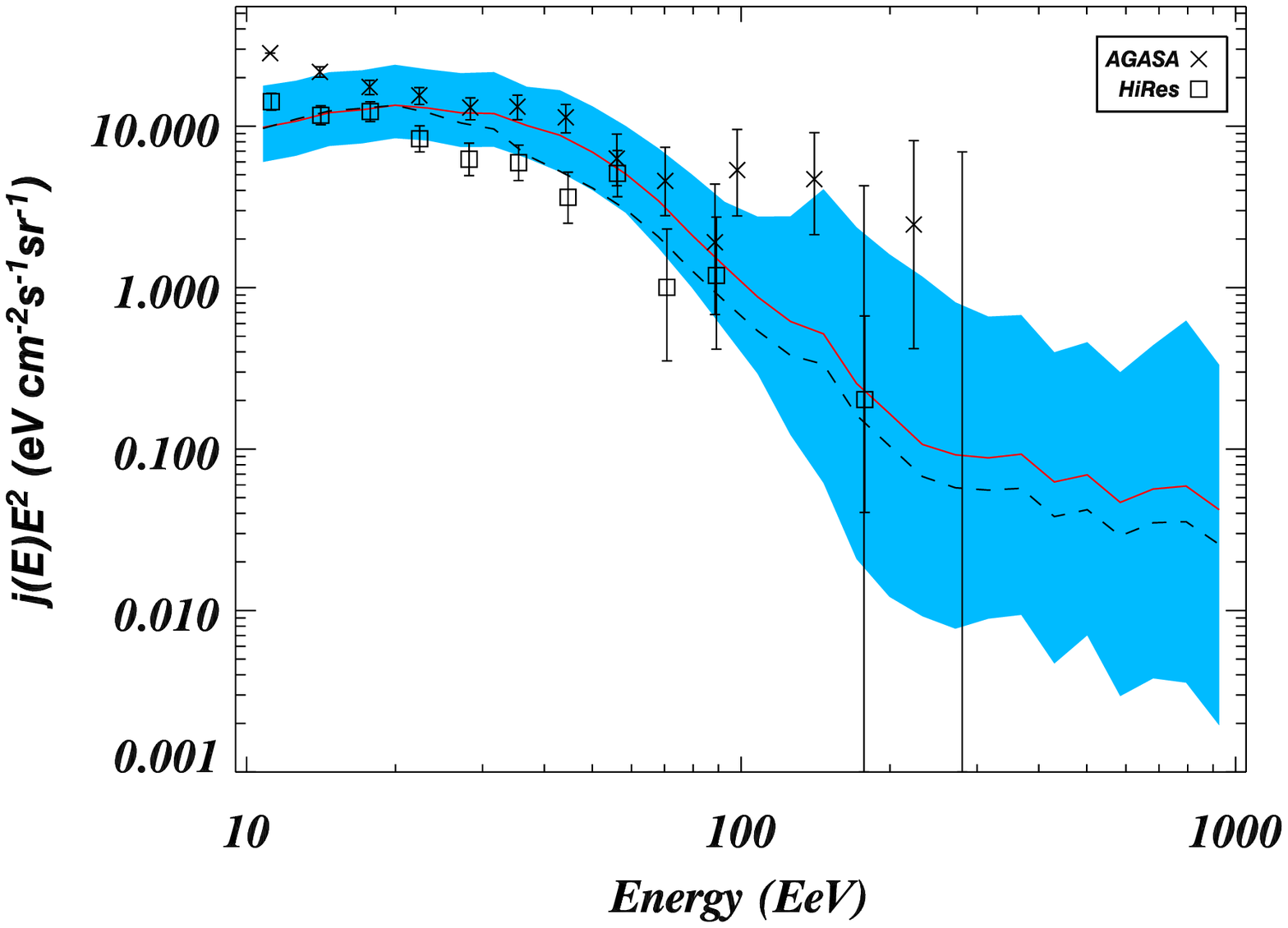}
\caption{Energy spectra predicted for iron injection in the presence
of the EGMF obtained from the LSS simulation. The average (solid line)
and cosmic variance (shaded band) are obtained as in
Figs.~\ref{fig:pspectrum_bfield} and~\ref{fig:iron_spectrum_nob}. %, based on
%50 source location realizations with 20 source property realizations each.
The average source spectral index is $\langle\alpha\rangle = 2.0$.
The maximum injection energy is set to 4 ZeV (top panel) and 10 ZeV
(bottom panel). The dashed lines represent average spectra in the case
$\langle \alpha \rangle = 2.4$. }
\label{fig:iron_spectrum}
\end{figure}

\begin{figure}[ht]
\includegraphics[width=0.48\textwidth,clip=true]{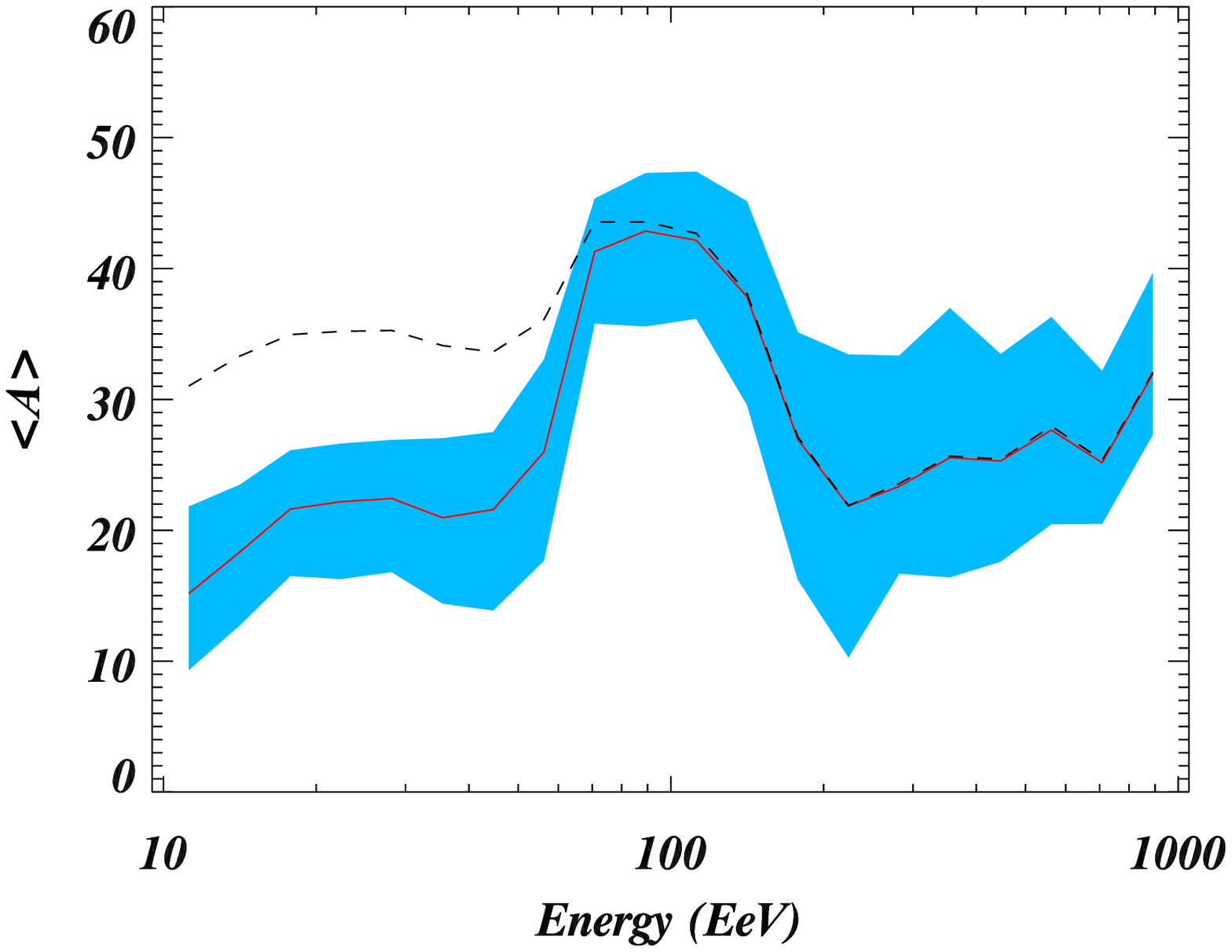}
\includegraphics[width=0.48\textwidth,clip=true]{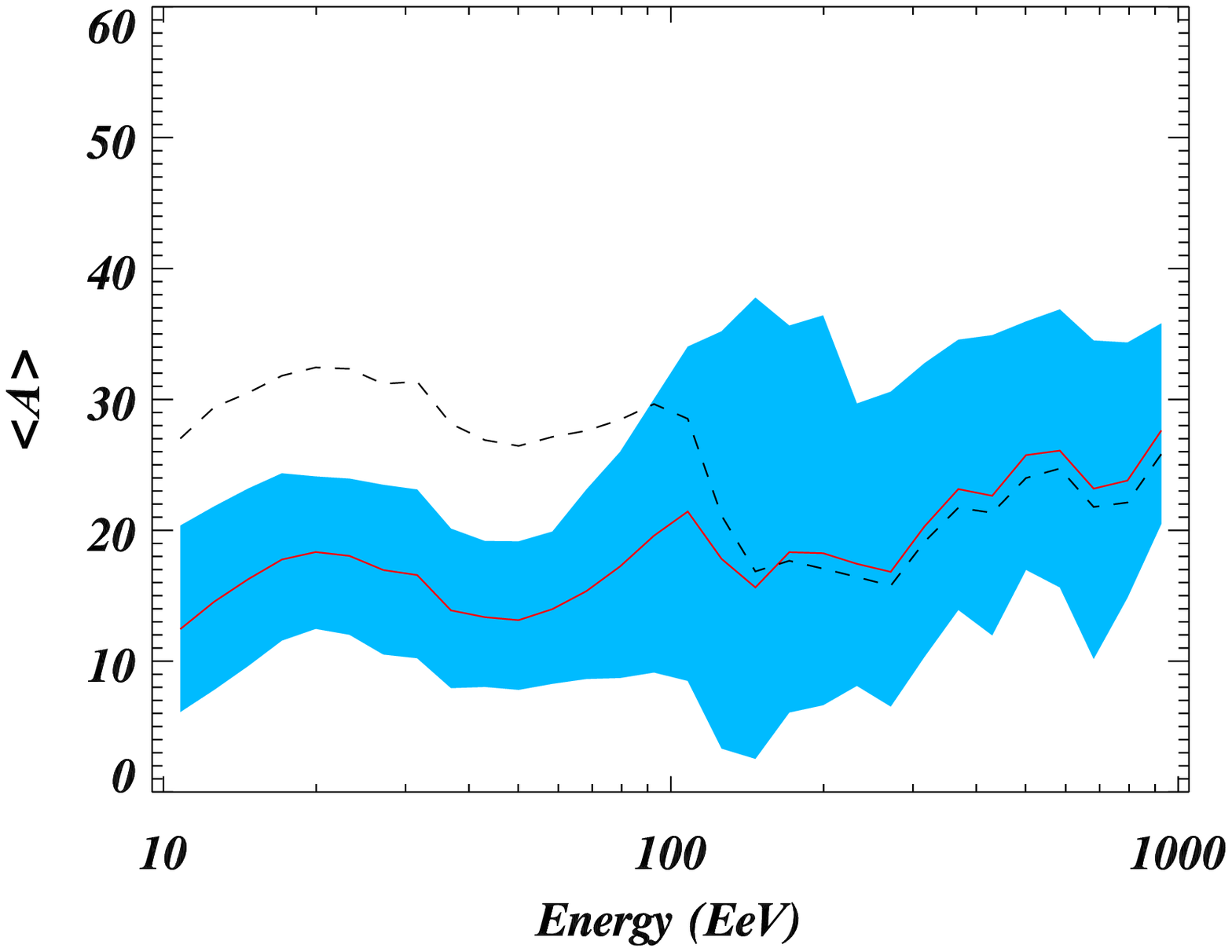}
\caption{Predicted average atomic mass $\langle A\rangle$ as a function
of observed particle energy corresponding to the spectra shown
in Fig.~\ref{fig:iron_spectrum} for iron injection in the presence of
the EGMF. Mean and cosmic variance are computed as in
Fig.~\ref{fig:iron_spectrum} with $\langle\alpha\rangle = 2.0$ and
maximum injection energy of 4 ZeV (top panel) and 10 ZeV (bottom panel).
The influence of the EGMF is apparent from comparing with
Fig.~\ref{fig:iron_compo_nob} for the same case without EGMF. The
dashed line corresponds to a mean injection
spectrum $\langle\alpha\rangle = 2.4$ instead of 2.}
\label{fig:iron_compo}
\end{figure}

The spectra obtained in presence of magnetic deflections are presented
in Fig.~\ref{fig:iron_spectrum}, and can be compared to
Fig.~\ref{fig:iron_spectrum_nob}. The bump which
was predicted between 20 and 30 EeV without EGMF now appears as a smooth
feature in a broader energy range, $10\lesssim E\lesssim 40\,$EeV. There
is no steepening of the flux at 100 EeV, comparable to
proton injection, and contrary to the case with no EGMF.

At higher energies, we observe a flattening of the spectrum as in the
absence of EGMF. The flux is higher than without EGMF, at
least statistically: For $E_{\rm max} = 4$ ZeV, we predict $\langle J
E^2\rangle \sim 0.1\,{\rm eV}\,{\rm cm}^{-2}\,{\rm s}^{-1}\,{\rm sr}^{-1}$
at $\sim 500$ EeV with EGMF, instead of
$\sim 0.01\,{\rm eV}\,{\rm cm}^{-2}\,{\rm s}^{-1}\,{\rm sr}^{-1}$
in the absence of EGMF.

The average atomic mass $\langle A\rangle$ as a function of energy
is presented in Fig.~\ref{fig:iron_compo}. At energies $E\lesssim60\,$EeV,
it can be seen that, depending on the average injection
spectral index $\langle \alpha \rangle$, $\langle A \rangle\sim 15-30$
instead of $\langle  A \rangle \sim 30-50$ in the absence of EGMF.
The interpretation is that, as explained in Ref.~\cite{bertone},
magnetic fields increase the mean path length between sources and
observer and the resulting increase in interactions also drives up
the relative proportion of lighter secondaries
to iron. As a consequence, it is very interesting to remark that
a measured average mass of UHECRs at energies $10\lesssim E\lesssim30\,$EeV
larger than $\simeq 35$ would imply that deflections due to EGMF
are relatively small, independently of the nature of
particles accelerated at the source. This can be another test for
extragalactic magnetic field effects, independently from anisotropy
studies~\cite{lss-protons}. The trend of $\langle A\rangle$ with
spectral index $\langle\alpha\rangle$ is the same as in the absence
of EGMF. Furthermore, the shape of the mass distribution is similar to
Fig.~\ref{fig:mass_histo}, except that abundances of
light elements with $A\lesssim6$ tend to be increased by a factor
2--3 which corresponds to the decrease in $\langle A\rangle$ seen
in comparing Fig.~\ref{fig:iron_compo_nob} with
Fig.~\ref{fig:iron_compo}.

At higher energies the energy dependence of $\langle A \rangle$
around $\sim100\,$EeV depends crucially on the source parameters
$E_{\rm max}$ and $\alpha$, as in the absence of EGMF.
The EGMF can lead to additional modification of the composition, as
seen by comparing Figs.~\ref{fig:iron_compo} with~\ref{fig:iron_compo_nob}.

%%%%%%%%%%%%%%%%%%%%%%%%%%%%%%%%%%%%%%%%%%%%%%%%%%%%%%%%%%%%%%%%%%%%%%%%%%

\subsection{Anisotropies: Deflections and Sky Maps}

\begin{figure}[!ht]
\includegraphics[width=0.48\textwidth,clip=true]{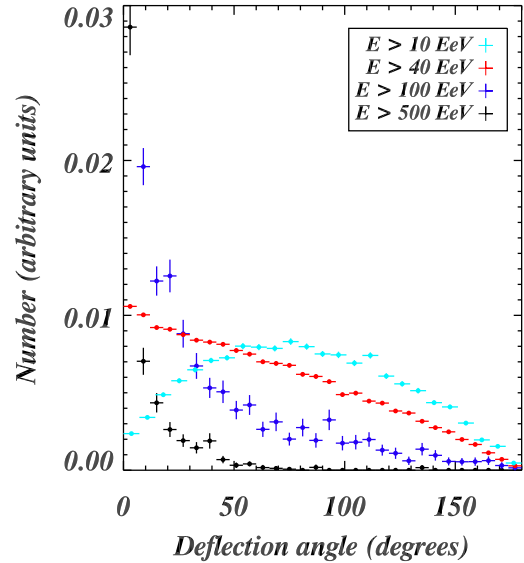}
\caption{Histogram of deflection angles of observed UHECRs for
different cuts in energy, in the case of iron injection with $\alpha =
2$. As in Fig.~\ref{fig:p_deflec}, the distributions are cumulated
over all source location realizations and the error bars are Poissonian,
reflecting the finite number of simulated trajectories.}
\label{fig:iron_deflec}
\end{figure}

\begin{figure}[h]
\includegraphics[width=0.28\textwidth,clip=true,angle=90]
{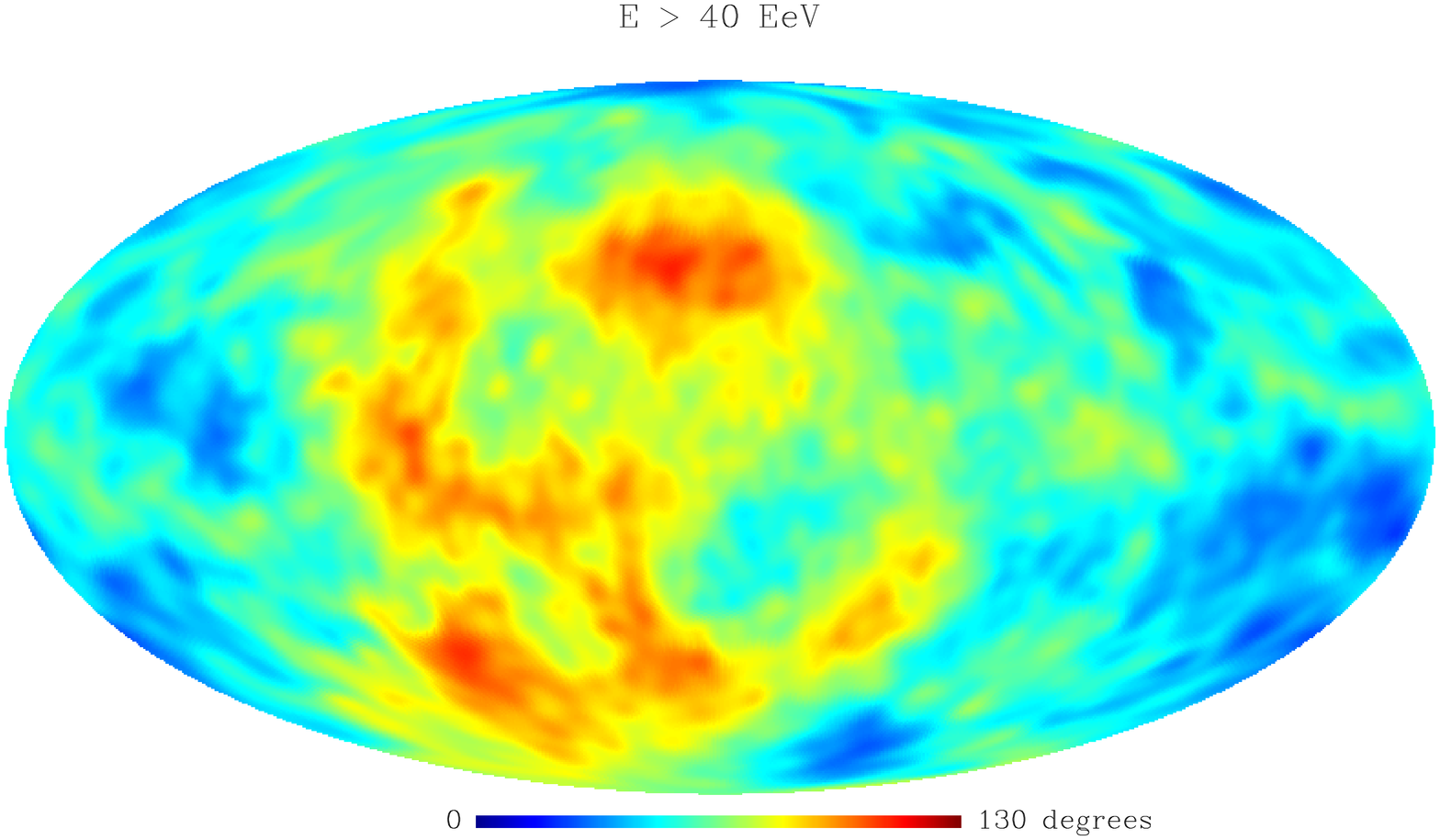}
\includegraphics[width=0.28\textwidth,clip=true,angle=90]
{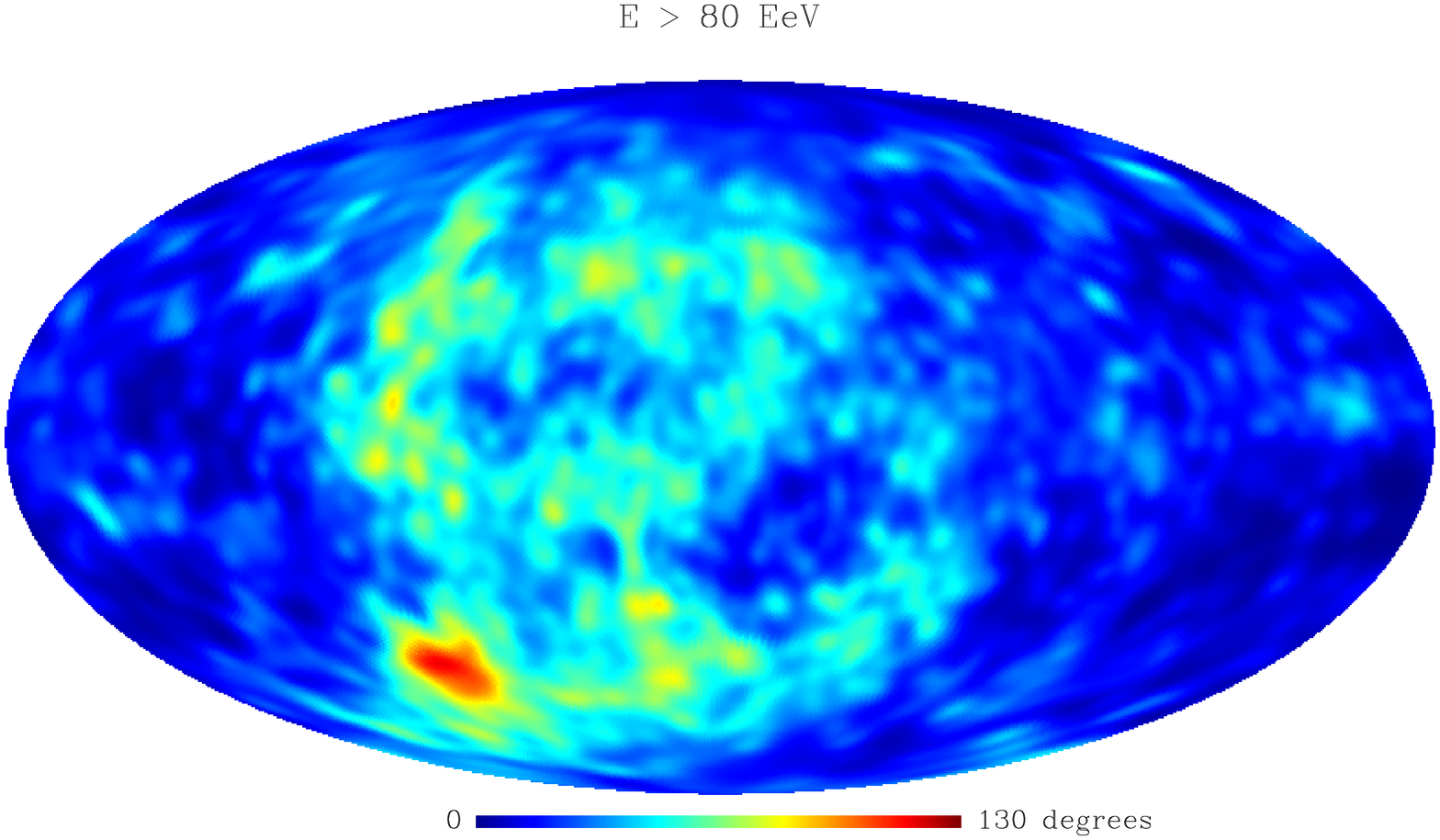}
\caption{Sky distribution of deflections obtained from events recorded by the
observer in the case of iron injection with EGMF, accumulated over all
simulated realizations. Top panel:
Events above 40 EeV. Bottom panel: Events above 80 EeV. The maps
have been smoothed with beams of 5 degrees.}
\label{fig:deflec_map}
\end{figure}

\begin{figure}[ht]
\includegraphics[width=0.28\textwidth,clip=true,angle=90]
{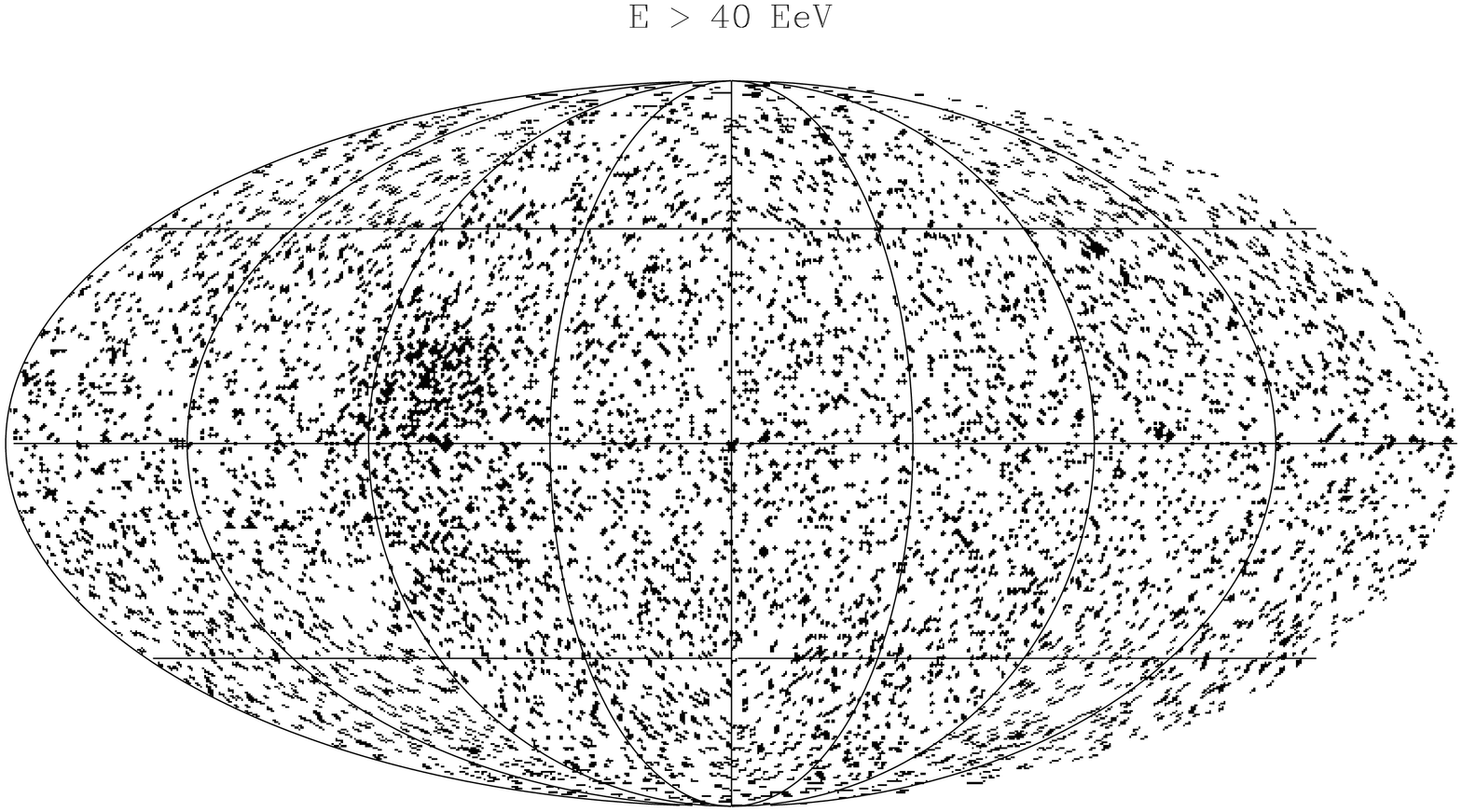}
\includegraphics[width=0.28\textwidth,clip=true,angle=90]
{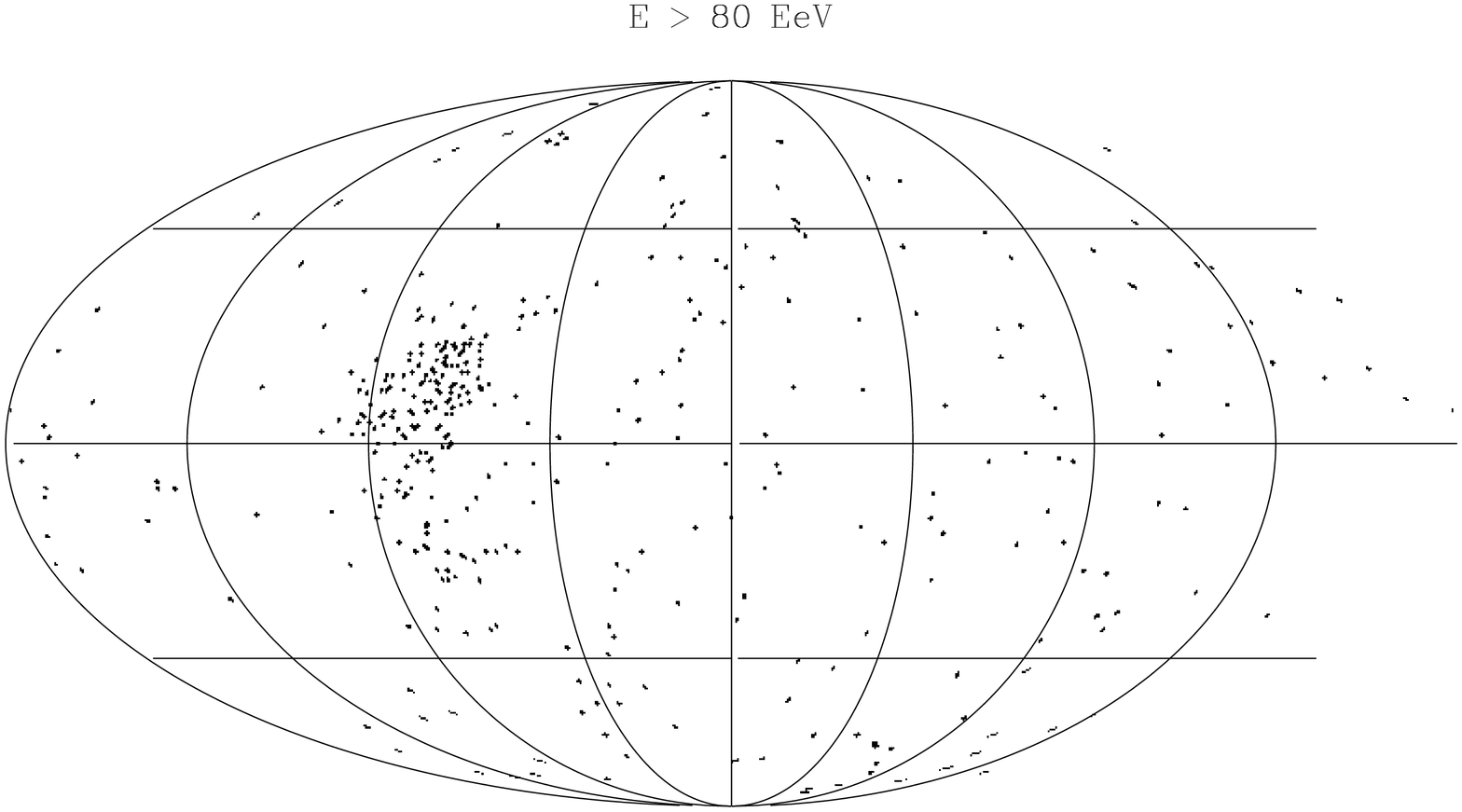}
\includegraphics[width=0.28\textwidth,clip=true,angle=90]
{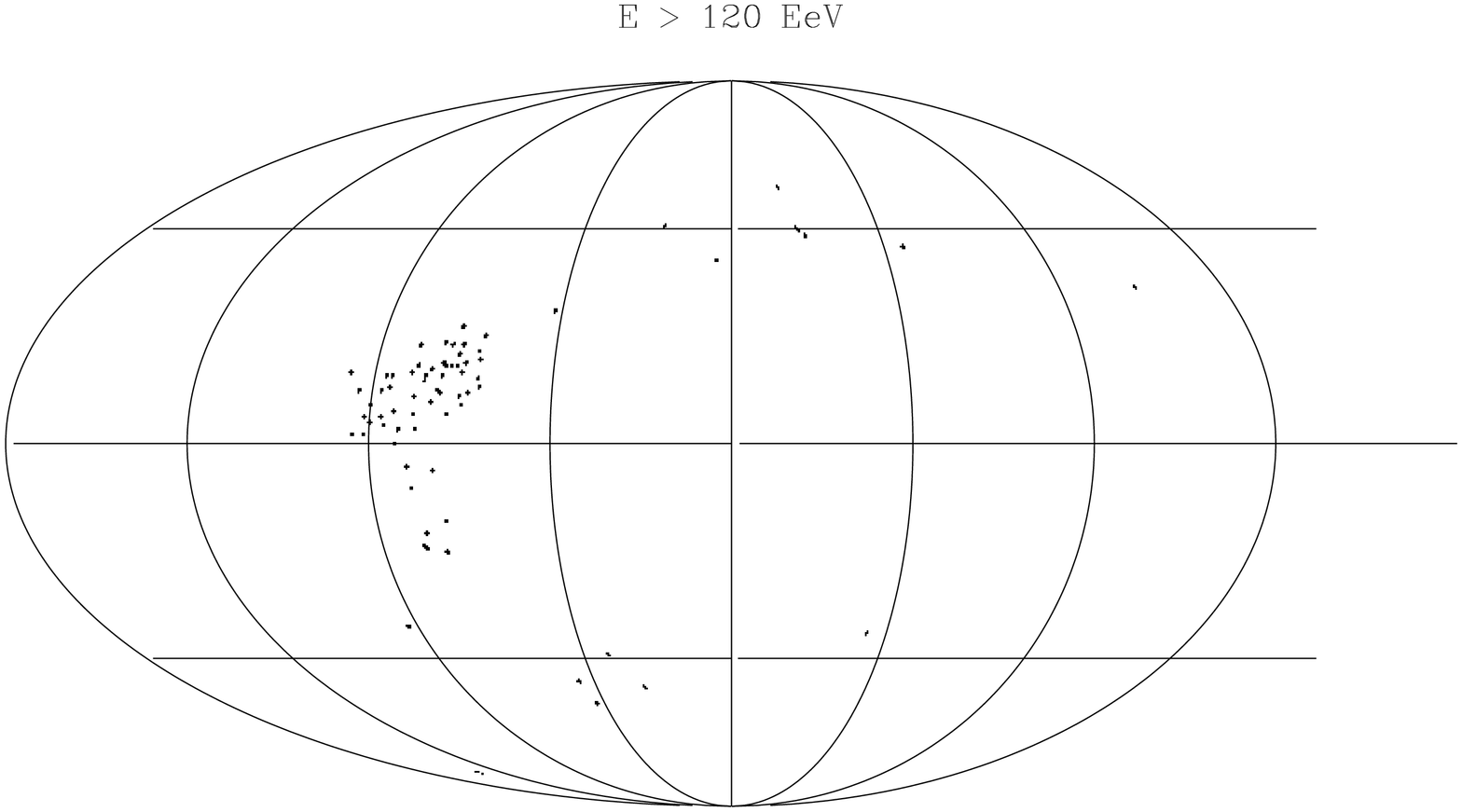}
\caption{Example of arrival directions of events above 40, 80, and 120 EeV
in a particular source realization,
with roughly 150 events recorded above 100 EeV. In this realization, at high
energy an individual source emerges progressively from the background.}
\label{fig:map_example_real}
\end{figure}

As expected, the histogram of deflection angles for iron injection,
presented in Fig.~\ref{fig:iron_deflec}, shows typical deflections
larger than for proton injection. This is particularly the case at the highest
energies, since at those energies the observed particles are mostly
heavy nuclei, whose local deflection is therefore up to 56 times larger
than in the case of protons. As a result, typical deflections of
$10^{\circ}$ or more are expected even at 500 EeV.

Sky distributions of deflection angles in the framework of this
scenario are presented in Fig.~\ref{fig:deflec_map} for energies
above 40 and 80 EeV. The considerable inhomogeneity of these maps
reflects the presence of extended magnetized structures distributed
along the LSS. There is a
striking difference between these maps and the one presented in
Ref.~\cite{dolag}: Even at 80 EeV, there is still a large part of the
sky where deflections are typically $\geq 50^{\circ}$.
This is due to at least two
reasons: Apart from the larger charge of the UHECRs in this scenario,
the EGMF in our simulation is more extended, as discussed previously.

These deflections may prevent us from performing straightforward
``UHECR astronomy'' with forecoming experiments such as the Pierre
Auger Observatories. However, they do not erase all the structures
in the sky either, and in particular the expected low source density
may allow to identify extended sources in the sky even within this
unfavorable scenario. An example is given in
Fig.~\ref{fig:map_example_real}, where arrival direction maps are
represented for a given source realization with iron injection and
EGMF, and a statistics of $\simeq 150$ recorded trajectories above
$10^{20}$ eV. At 40 EeV, the sky is dominated by an isotropic cosmological
background, but at higher energies the background disappears and a
small number of nearby sources emerge. These sources are clearly
visible although they
are not point-like. It has to be noticed that, if such a configuration
were realized in nature, then the very few sources one expects to
contribute at the highest energies can be found in the northern
hemisphere, where Virgo is located, emphasizing the need for a
northern UHECR observatory.

%%%%%%%%%%%%%%%%%%%%%%%%%%%%%%%%%%%%%%%%%%%%%%%%%%%%%%%%%%%%%%%%%%%%%%%%%%

\subsection{Anisotropies: Auto-Correlation and Angular Power Spectrum}

\subsubsection{Auto-Correlation Function}

\begin{figure}[!ht]
\includegraphics[width=0.48\textwidth,clip=true]
{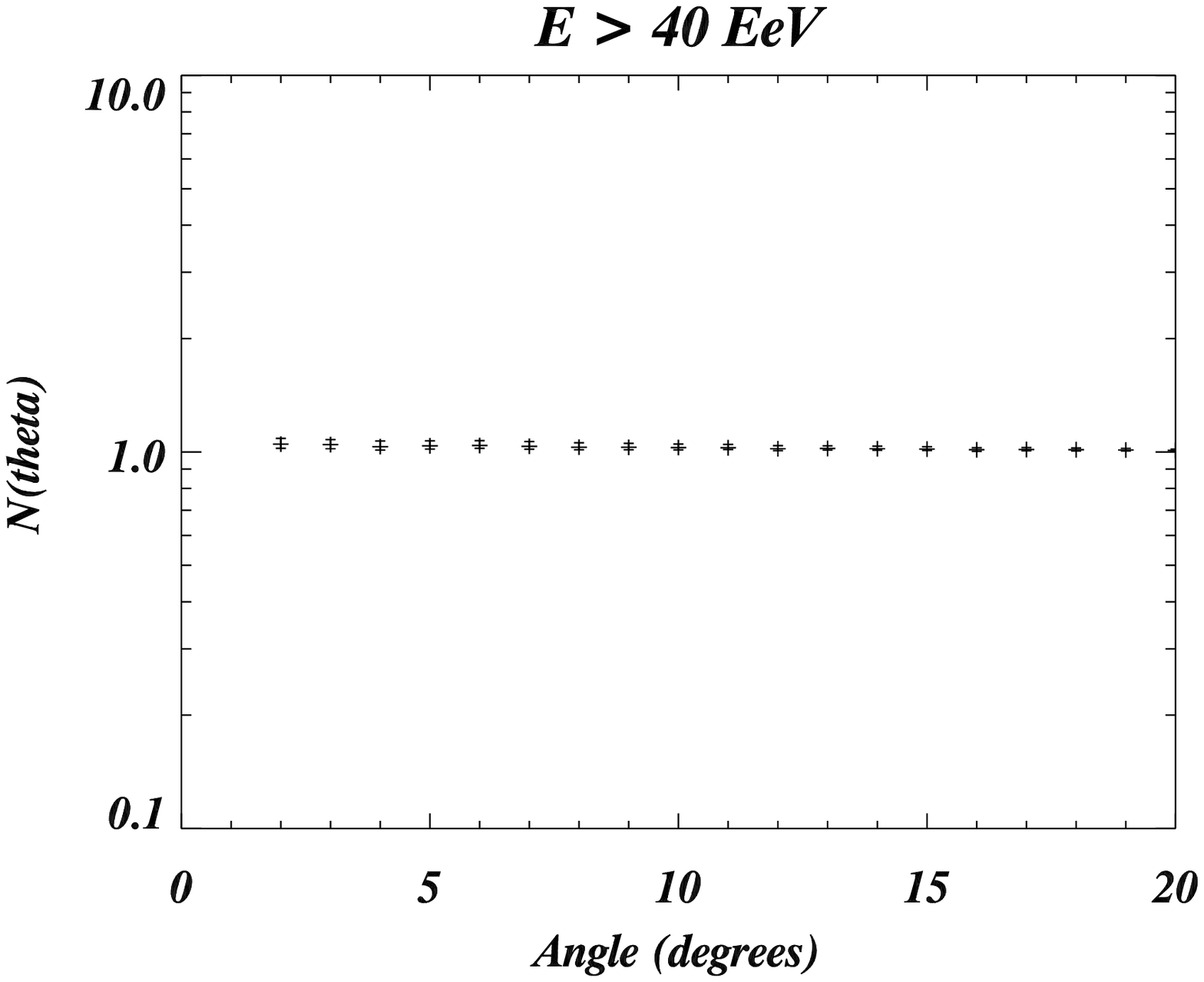}
\includegraphics[width=0.48\textwidth,clip=true]
{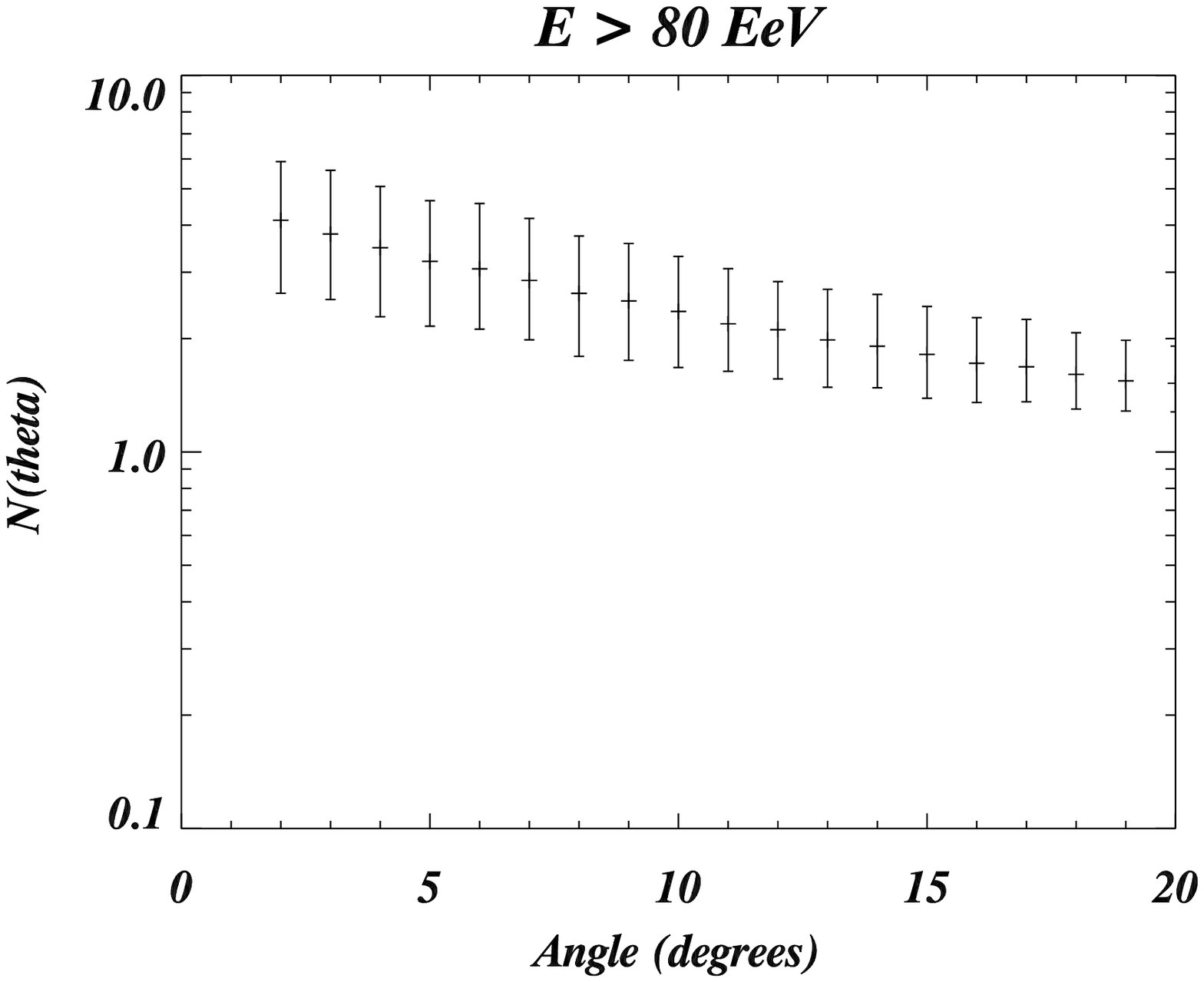}
\includegraphics[width=0.48\textwidth,clip=true]
{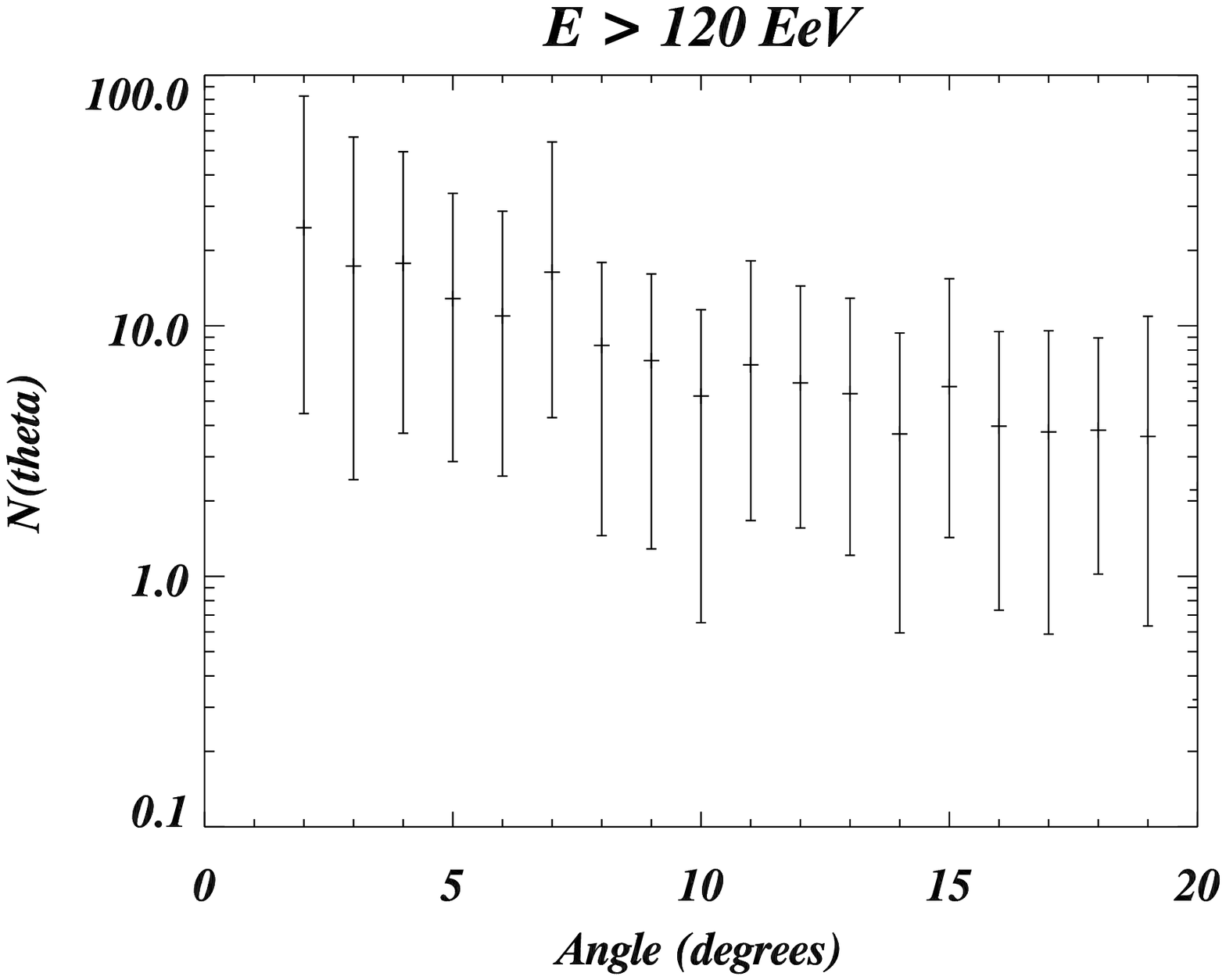}
\caption{Auto-correlation function for events above 40, 80 and 120 EeV
in the case of iron injection with EGMF. We do not
show the auto-correlation in the first bin and showers have been
corrected for. The error bars represent the cosmic variance, computed in
the same way as for spectra and composition: Various source location
configurations are considered, and for each one 50 source intensity
and spectral indexes are drawn. ${\cal N}(\theta) = 1$ corresponds to
isotropy.}
\label{fig:iron_autocor}
\end{figure}

The large deflections do not prevent us from observing a
strong auto-correlation signal at high energies, due to the low source
density. We present in Fig.~\ref{fig:iron_autocor} the auto-correlation
predicted for this scenario, computed with a 1 degree binning. Due to the
computational reasons discussed in Sect.~II, we count nuclei showers
as only one particle and we disregard the first angular bin. It can be
seen that above 40 EeV, the predicted auto-correlation signal is almost
flat, whereas above 80 EeV it reaches a factor $\sim 4$ above the
isotropic level, and much larger values above 120 EeV.
Fig.~\ref{fig:iron_autocor} also shows that the predicted
auto-correlation signal is highly dependent on the source configuration,
and can extend over 10 degrees or more due to diffusion in the
EGMF.

When comparing Fig.~\ref{fig:iron_autocor} to
Fig.~\ref{fig:iron_autocor_nob}, it can be seen that the effect of
magnetic fields is to strongly reduce the intensity of the
auto-correlation peak at small angles: at 80 EeV, the reduction
factor is $\sim 10$. Due to the larger deflection of nuclei this
suppression is stronger than for protons.

We remark that the galactic magnetic fields, which were not
take into account here, will also increase
this smoothing of auto-correlation over larger angular scales.

%%%%%%%%%%%%%%%%%%%%%%%%%%%%%%%%%%%%%%%%%%%%%%%%%%%%%%%%%%%%%%%%%%%%%%%%%%

\subsubsection{Angular Power Spectrum}

\begin{figure}[!ht]
\includegraphics[width=0.48\textwidth,clip=true]{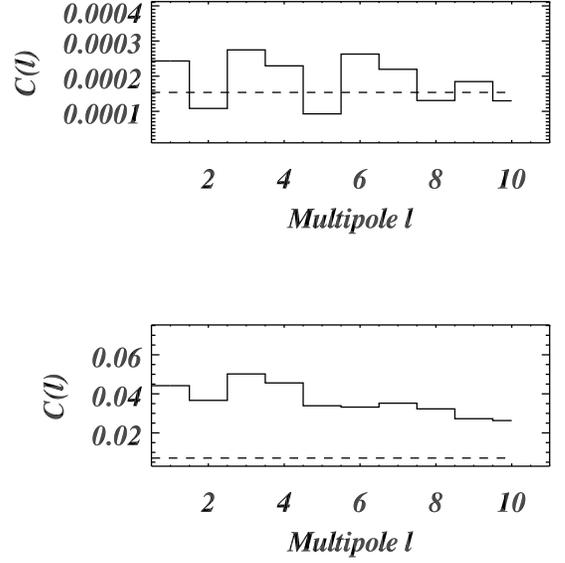}
\caption{Example of predicted angular power spectra for two source
configurations for the scenario with iron injection and EGMF.
Top panel: Events with $E\geq40\,$EeV; the corresponding event map is
roughly isotropic and contains several thousand events. With reweighting,
we predict $\left< C_{\ell} \right> = 1.54\times10^{-4}$, i.e. $N_{\rm
eff} \simeq 500$ effective events, see Eq.~(\ref{c2}); this level
is indicated by the dashed line.
Bottom panel: Events with $E \geq 120\,$EeV. Two nearby sources
appear in the corresponding map. The average $\left< C_{\ell} \right>$
predicted from isotropy is $7.1\times10^{-3}$ (dashed line),
and the average $\left< C_{\ell} \right>$ predicted by the simulation
implies $N_v=2.2$ ``effective point sources''. Furthermore, $C_{\ell}$
decreases with increasing $l$, which reflects the spatial extension
of the sources.}
\label{fig:cl_example}
\end{figure}

\begin{figure}[!ht]
\includegraphics[width=0.48\textwidth,clip=true]
{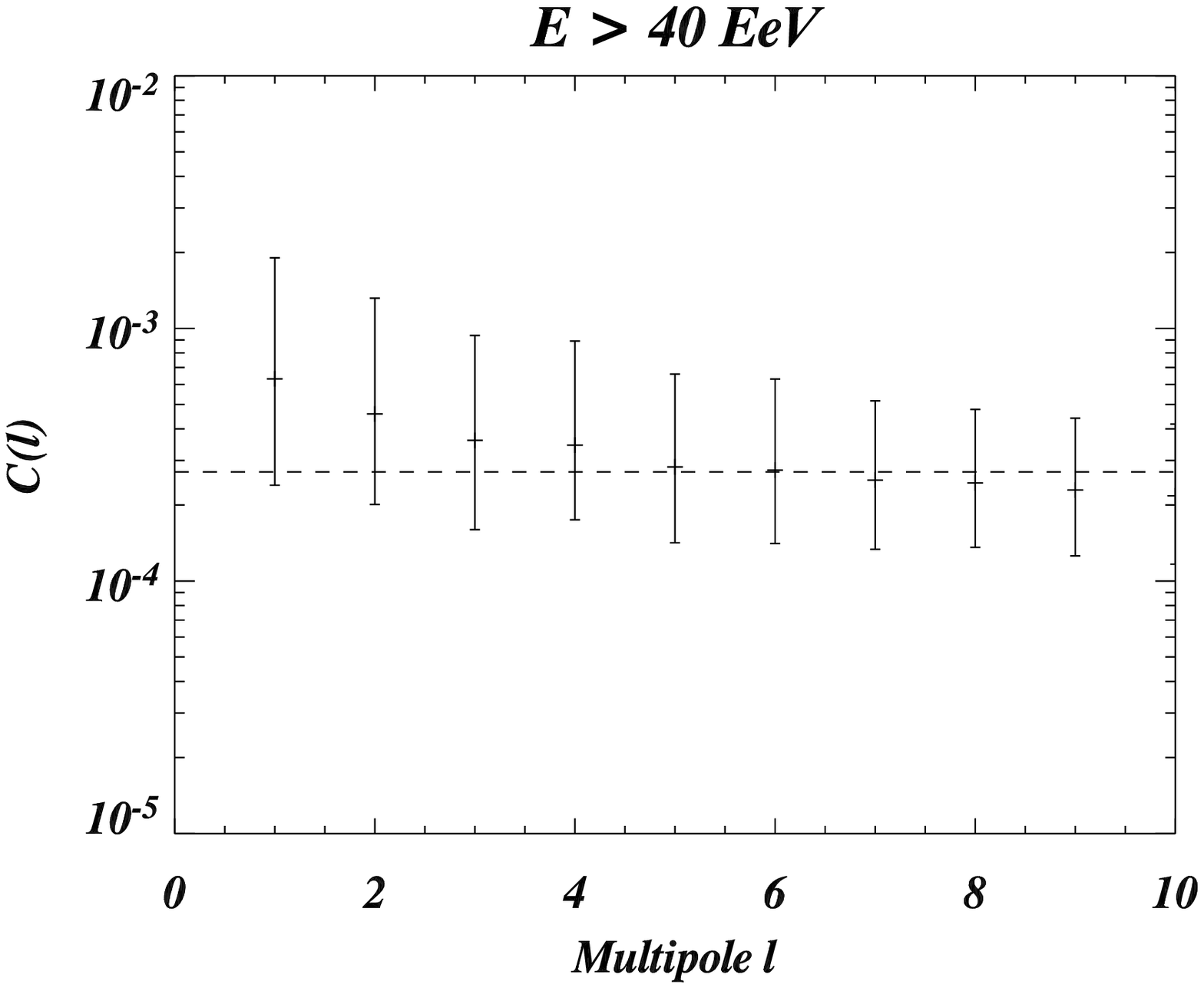}
\includegraphics[width=0.48\textwidth,clip=true]
{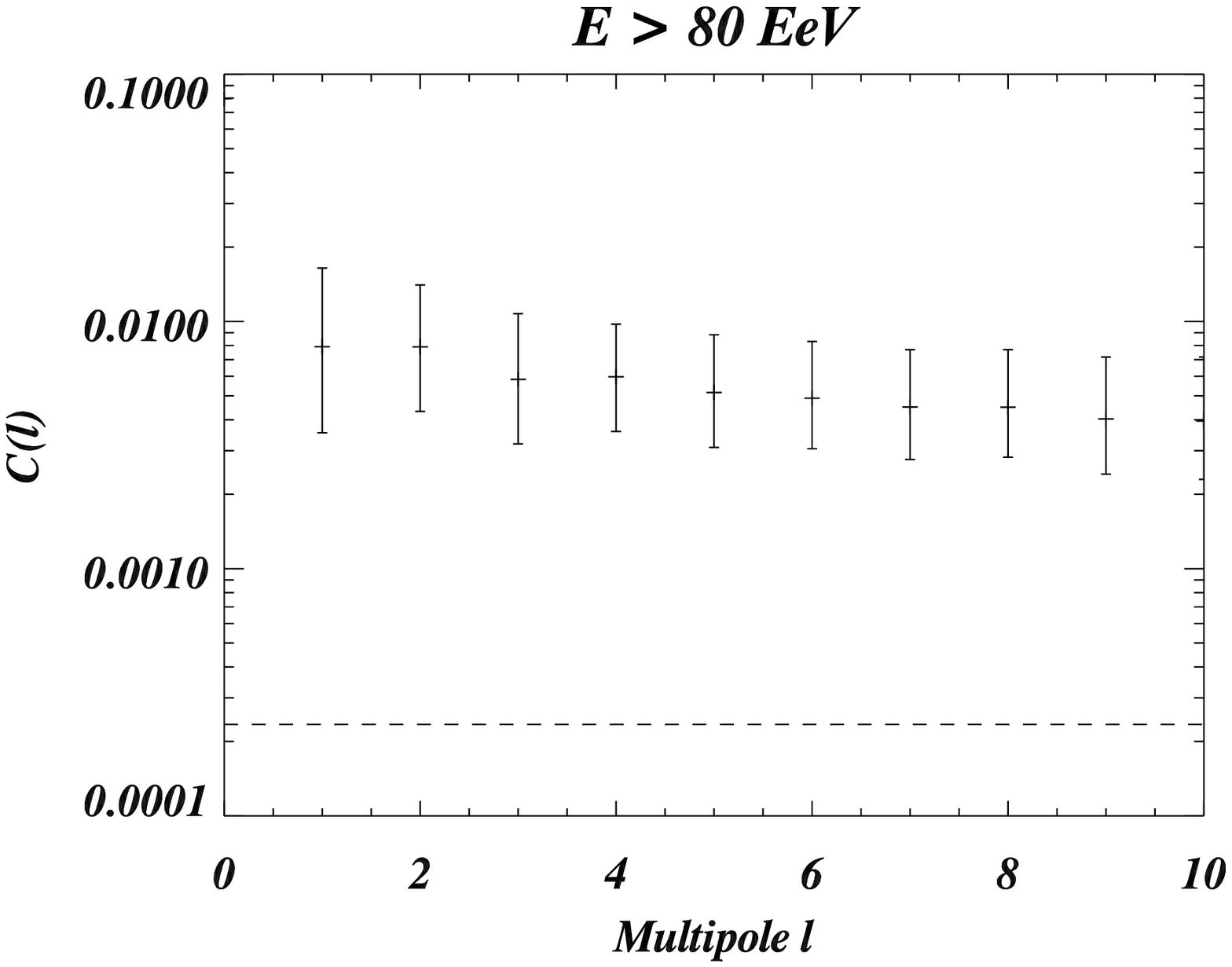}
\includegraphics[width=0.48\textwidth,clip=true]
{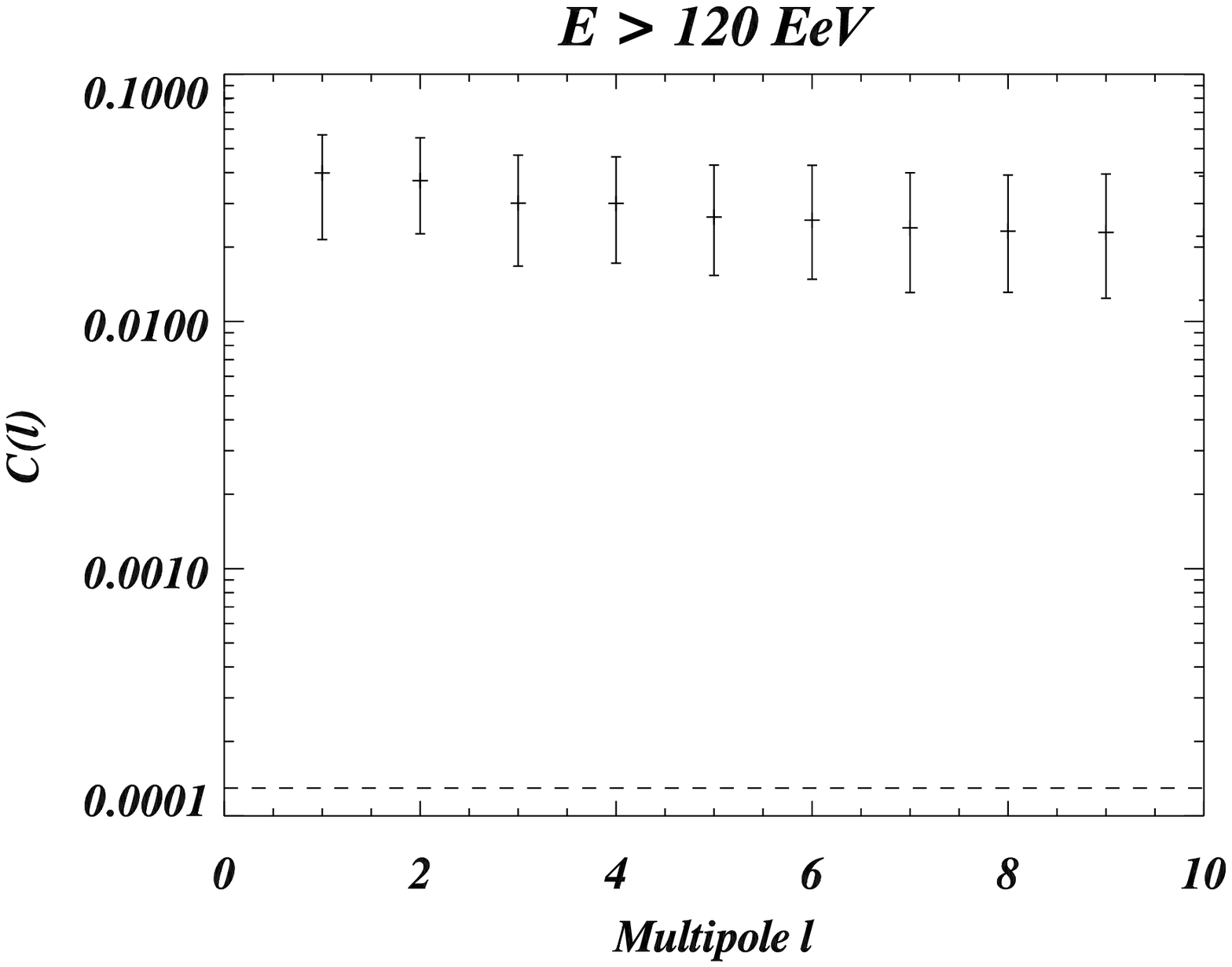}
\caption{Angular power spectra for events above 40, 80 and 120 EeV
in the case of iron injection with EGMF. The error bars
represent the cosmic variance computed in the same way
as in Fig.~\ref{fig:iron_autocor}. The dashed line corresponds
to an isotropic distribution of the same number of $10^4$ simulated,
weighted events above 40 EeV.}
\label{fig:iron_cl}
\end{figure}

In Figs.~\ref{fig:cl_example} and~\ref{fig:iron_cl}, we present
examples of angular power spectra, and their averages and fluctuations
for given energy thresholds. The predictions are similar to those
obtained in the absence of EGMF in Fig.~\ref{fig:iron_cl_nob}.
The theoretical prediction for the power spectrum in the isotropic
case is given by the expressions in Eq.~(\ref{c2}).
As the large fluctuations show, the power spectrum depends on
the energy threshold, as well as on the source configuration.

Above 40 EeV the mean angular power spectrum is almost flat and
consistent with predictions from isotropy; however some realizations
exhibit some large-scale patterns visible in the low-order
$C_{\ell}$. When increasing the energy, the angular power spectrum remains
roughly flat, but at levels much higher than expected from
isotropy. Indeed, at high energies there is only a small number
$N_v$ of bright sources visible in any given realization, leading to
a mean value of $C_{\ell} = (4\pi N_v)^{-1}$. The power spectrum is then
approximatively flat, with deviations from flatness caused by the
observed source extension due to deflection in the EGMF. Above 80 EeV,
the average $\langle C_{\ell} \rangle$ is
$5\times10^{-3}$, leading to an average value $N_v \simeq 16$ visible sources,
whereas above 120 EeV we have $\left< C_{\ell} \right> \sim 0.03$, and
therefore, only 2.7 sources are visible on average.

The angular power spectrum is a useful tool to estimate the
effective number of observed sources above a given energy, but
otherwise it appears to be of quite limited use within the scenario under
consideration: At energies below 40 EeV, deflections are too
large to expect a large-scale signal, and at high energies the
large values predicted for $C_{\ell}$ are only due to the finite number
of visible sources.

%%%%%%%%%%%%%%%%%%%%%%%%%%%%%%%%%%%%%%%%%%%%%%%%%%%%%%%%%%%%%%%%%%%%%%%%%%

\subsubsection{Detection with Finite Statistics}

When comparing predictions of auto-correlations and power spectra such
as shown in Figs.~\ref{fig:iron_autocor}--\ref{fig:iron_cl} with future
data, one has to take into account
also the statistical error due to the finite number $N_{\rm obs}$ of
the actually observed events. This has not been included in
Figs.~\ref{fig:iron_autocor}--\ref{fig:iron_cl} which thus correspond
to $N_{\rm obs} \rightarrow \infty$.
The error bars due to finite statistics can be assigned directly to
the observational angular power spectra or auto-correlations,
as these errors depend on the detector aperture.

In the same way, we have not taken into account the detector exposure
in our anisotropy predictions, as it can be taken into account directly in
the experimental data.

Alternatively, the finite number of experimental events and
non-trivial exposure functions as well as angular and energy
resolutions can be assigned to the predictions by
drawing mock data sets from maps constructed from the simulated
trajectories, as done in Ref.~\cite{lss-protons}.

However, we can roughly estimate the number of events needed in the
case of uniform exposure to detect the anisotropies predicted in our
model. This
can be done by estimating the error due to the finite number of events
and requiring it to be smaller than cosmic variance and/or the
deviation from predictions based on isotropy.

Even in the case of partial sky coverage, the statistical error on the
angular power spectrum estimation has been computed~\cite{deligny},
but the isotropic formula $\sigma(C_{\ell}) \sim
C_{\ell} \simeq 1/(4\pi N_{\rm obs})$ ($C_{\ell}$ is a quadratic estimator),
at low $\ell$, is sufficient for our purposes. From Fig.~\ref{fig:iron_cl}, we
predict $C_{\ell} \sim 0.005$ above 80 EeV. Requiring
$\sigma(C_{\ell}) \sim 0.001$ for a detection implies to have
$N_{\rm obs}\gtrsim 80$ events above $8\times10^{19}$ eV, which
has not been yet reached by AGASA, but will be easily reached by
the Pierre Auger project.

As for the auto-correlation function ${\cal N}(\theta)$, we predict
an excess $\cal N$~$ \sim 4$ for $\theta \lesssim
\theta_0 = 2^{\circ}$ at 80 EeV, see Fig.~\ref{fig:iron_autocor}. For
$N_{\rm obs}$ events, the Poissonian error on
$\cal N$ is $\sigma({\cal N}) \simeq(N_{\rm obs}^2\theta_0^2
/4)^{-1/2}$. Requiring $\sigma({\cal N}) \leq 0.8$ for a detection leads to
$N_{\rm obs}\gtrsim 70$ events needed above $8 \times 10^{19}$ eV.

Overall, the anisotropies arising in the framework of this model are
largely detectable by the Pierre Auger Observatory.

%%%%%%%%%%%%%%%%%%%%%%%%%%%%%%%%%%%%%%%%%%%%%%%%%%%%%%%%%%%%%%%%%%%%%%%%%%

\section{Conclusions}

We performed numerical simulations of ultrahigh energy nuclei
propagation in a structured, magnetized universe.
The sources, of density $\sim 10^{-5} \mbox{Mpc}^{-3}$,
were distributed proportionally to the baryon density obtained
from the large scale structure simulation of Ref~\cite{miniati}. We
considered either negligible magnetic fields or fields obtained
from that same simulation of large scale structure formation
and seeded by the Biermann battery around cosmic shocks.
Their strength has been normalized to $\sim \mu$G in the core of
the most prominent simulated collapsed structure~\cite{lss-protons}.
We injected iron and protons at the sources, and in the case
of iron we followed all secondary nuclei produced by photo-disintegration.

Below the GZK feature, protons tend to dominate the spectrum
even for exclusive iron injection, due to photo-disintegration of
high-energy heavy nuclei. However, the predicted spectrum does not match
the observations by AGASA and HiRes at these energies, and a light
component from injection of light nuclei or protons must contribute
to the spectrum at sub-GZK energies.

Above $\simeq30\,$EeV existing data are poor and consistent with an UHECR
flux exclusively resulting from iron injection. As the maximum
energy reachable with shock acceleration increases with the charge
of the nucleus, it makes sense to expect a heavy nuclei component
above the GZK energy. Within our scenario we found the following
generic predictions in the case of pure iron injection:

\begin{itemize}
\item{The average mass $\langle A \rangle$ of observed particles
strongly depends on the scenario and typically is $\sim 20 - 30$ above
$\simeq 100$ EeV. If
$\langle A \rangle\gtrsim35$ is observed at energies below 30 EeV, the
effect of extra-galactic
magnetic fields on propagation cannot be too strong since it would
increase photo-disintegration. Below $\simeq100\,$EeV, $\langle A
\rangle$ increases with increasingly steeper injection spectrum.}
\item{If some sources are located within a few Mpc from the observer,
a flattening
of the spectrum after the GZK feature results. This flattening
is more pronounced than in the case of proton
primaries, but cosmic variance is much larger as well. When the
spectrum is normalized around 30 EeV, the flattening
tends to be stronger for significant EGMF.}
\item{A low source density and considerable magnetic fields,
especially around the sources, predict a significant clustering
of UHECRs at super-GZK energies, typically on scales of
order of the angular size of the magnetized region around the
sources. In our simulation it is of order
$\sim 10^{\circ}$. This can be seen from both auto-correlation
function and angular power spectrum measurements at various energies.}
\item{The injection spectra which
allow to fit the observed spectra tend to be somewhat
harder than in the case of protons, and cover the range
$2.0\lesssim\alpha\lesssim2.4$, but still consistent
with expectations from theory of shock acceleration~\cite{shock-acc}.}
\end{itemize}

The main feature of these simulations compared to previous work is
the distribution of structured magnetic fields, compatible with
existing data on extragalactic fields. These fields increase the
deflection of nuclei, and, as a consequence, smooth the GZK feature in
the energy spectrum as well as the auto-correlation signal at small angles.
We also found that, for iron injection, there is a major cosmic
variance at the highest energies for observables such as energy
spectrum, composition and angular distribution, mostly due to the
uncertainties in the location of the sources.

At the highest energies the observed flux will be dominated by
only a few sources within the GZK-distance, $\simeq50\,$Mpc,
from the observer. Although extra-galactic magnetic fields can
considerably smear out the images of the sources, their location
may still be determined by next generation experiments to within
$10-20^\circ$.

In the presence of EGMF auto-correlations are reduced by factors
of order 10, and thus more strongly than in the case of
proton injection~\cite{lss-protons}. However, even in the presence of
the relatively strong EGMF considered in the present work,
auto-correlations remain significant. As a consequence, if no
auto-correlation is observed, the source density must be larger
than $10^{-5} \mbox{Mpc}^{-5}$. Recall that this density is motivated
by the claimed AGASA clustering~\cite{blasi,kachelriess}.

We have now explored an extensive series of plausible scenarios in the
context of an astrophysical UHECR origin. The scenario mostly
studied here, because of the large magnetic fields involved over
extended regions, of the small source density compared to interaction
lengths, and of the fact that we inject iron nuclei at the sources,
appears to be the ``worst case'' in terms of fluctuations of
the predictions of observables. Hopefully, the
Pierre Auger Observatory will soon allow to select a specific scenario
if its data is compatible with an astrophysical origin of UHECRs.

\section*{Acknowledgments}
Maps are pixellized and projected using Healpix softwares~\cite{healpix}.
FM acknowledges partial support by the Research and
Training Network ``The Physics of the Intergalactic Medium''
set up by the European Community
under the contract HPRN-CT2000-00126 RG29185
and by the Swiss Institute of Technology through a Zwicky
Prize Fellowship.

\end{document}